\documentclass[aps,prl,reprint,superscriptaddress,amsmath,amssymb]{revtex4-2}
\usepackage{color}
\usepackage{amsmath,amssymb}
\usepackage{graphicx}
\usepackage[colorlinks, linkcolor=blue, citecolor=blue, urlcolor=blue, breaklinks=true]{hyperref}
\usepackage{mathtools}

\DeclarePairedDelimiter\bra{\langle}{\rvert}
\DeclarePairedDelimiter\ket{\lvert}{\rangle}
\DeclarePairedDelimiterX\braket[2]{\langle}{\rangle}{#1\,\delimsize\vert\,\mathopen{}#2}

\begin{document}
\title{Displacement-noise-free gravitational-wave detector using quantum tomography of photon orbital angular momentum states}
\author{Haorong Wu}
\affiliation{School of Physics and Technology, Wuhan University, Wuhan 430072, China}
\author{Xilong Fan}
\email{xilong.fan@whu.edu.cn}
\affiliation{School of Physics and Technology, Wuhan University, Wuhan 430072, China}
\author{Lixiang Chen}
\email{chenlx@xmu.edu.cn}
\affiliation{Department of Physics, Xiamen University, Xiamen 361005, China}

\begin{abstract}	
	Gravitational waves (GWs) induce quantum transitions between photon orbital-angular-momentum (OAM) states with amplitudes $\alpha_{\pm}$. Exploiting this effect, we present a displacement-noise-free detection framework that extracts GW information from the OAM Hilbert space. Parametrized quantum state tomography of the GW-induced OAM sidebands enables a single V-shaped detector to resolve all seven wave parameters: the frequency, propagation direction, two polarization amplitudes, and initial phase of each polarization. Direct detection of the sidebands yields a weak signal of approximately $10^{-15}\,\mathrm{W}$. We therefore introduce a stronger reference state $r\ket{0}_{\mathrm{P}}$, with $r=0.01$, to enhance the detected signal. For an arm length of $L=4\,\mathrm{m}$, $N=160$ round trips, a paraxial parameter of $\gamma=0.1$, an incident light power of $P_{\mathrm{BS}}=10\,\mathrm{kW}$, a wavelength of $\lambda_{\mathrm{opt}}=1064\,\mathrm{nm}$, and GW amplitudes of $A=10^{-18}$, we estimate sideband transition amplitudes of $|\alpha_{\pm}|\sim10^{-10}$ and signal powers of order $1\,\mathrm{nW}$, scaling as $P_{\mathrm{BS}}\alpha_{\pm}\propto P_{\mathrm{BS}}ANL\gamma/\lambda_{\mathrm{opt}}$. The shot-noise-limited sensitivity floor is $S^{1/2}_n=1.6\times10^{-20}\,\mathrm{Hz}^{-1/2}$. Because the scheme measures OAM transition amplitudes rather than differential arm lengths, it is intrinsically free from displacement noise and provides a broadband probe of GWs, specifically bridging the $0.1$--$10\,\mathrm{Hz}$ mid-frequency gap.
\end{abstract}

\maketitle

% \section{Introduction}
\textit{Introduction}---Since the first direct detection of gravitational waves (GWs) by LIGO, the global detector network has recorded hundreds of signals \cite{abbott2016observation,PhysRevLett.119.161101,PhysRevX.13.041039}, whose source parameters are reconstructed using matched filtering and Bayesian inference \cite{maggiore2008gravitational}. Recently, we demonstrated that the Hilbert space spanned by photon orbital angular momentum (OAM) encodes GW information \cite{q38s-k2tq}
\begin{equation}
	h_{\mu\nu}(t,\mathbf x)=\sum_{\lambda} e_{\mu\nu}^\lambda(\theta,\phi) A_\lambda \cos(k_g (-t+\hat {\mathbf n}\cdot \mathbf x )+\Psi_\lambda),\label{eq.DefGW}
\end{equation}
where $\lambda=+,\times$ is the polarization, $e_{\mu\nu}^\lambda(\theta,\phi)$ are the unit polarization tensors and $\hat {\mathbf n}$ is the propagation direction, defined by the polar angle $\theta$ and azimuthal angle $\phi$. With $k_g$ representing the frequency, and $A_\lambda$ and $\Psi_\lambda$ denoting the polarization amplitudes and initial phases \cite{maggiore2008gravitational}, seven wave parameters ($k_g$, $\theta$, $\phi$, $A_\lambda$, and $\Psi_\lambda$) fully characterize the signal. However, Equation \eqref{eq.DefGW} shows that $k_g$ and $\Psi_\lambda$ are coupled with $A_\lambda$. This means measurements at any specific time $t$ only resolve the effective amplitudes $\tilde{A}_\lambda=A_\lambda \cos(k_g (-t+\hat {\mathbf n}\cdot \mathbf x )+\Psi_\lambda)$. Consequently, resolving the wave parameters $(\theta,\phi,\tilde{A}_+,\tilde{A}_\times)$ simultaneously requires four degrees of freedom (DOFs). Extracting these parameters within the OAM Hilbert space provides a novel framework for GW data analysis.

Since the foundational work of Allen \textit{et al.} \cite{allen1992orbital}, the spatial phase structure of photon OAM has been widely used in optics \cite{allen1992orbital,mair2001entanglement,vaziri2002experimental,vaziri2003concentration,molina2007twisted}, driving recent investigations at the intersection of OAM and GW physics. For instance, it has been demonstrated that higher-order Laguerre-Gaussian modes can reduce thermal noise in GW interferometers \cite{PhysRevD.79.122002,PhysRevLett.105.231102}. Furthermore, Wu \textit{et al.} established that the helicity coupling between GWs and structured light drives photons into different OAM eigenstates, yielding measurable optical signatures \cite{wu2022testing}. The work by Wu \textit{et al.} explored how the GW background degrades coherence and high-dimensional OAM entanglement as twisted photons propagate through curved spacetime \cite{PhysRevD.106.045023}.

Building on this, we previously demonstrated that passing GWs induce transitions in scalar and vector vortex beams from an initial OAM state $\ket{l}$ to sideband states $\ket{l\pm 1}$ and $\ket{l\pm 2}$ \cite{q38s-k2tq}. Because transitions to $\ket{l\pm 2}$ are strongly suppressed, the detectable signal is predominantly confined to the $\ket{l\pm 1}$ subspace. Most importantly, during the interaction, the GW information is encoded within the complex amplitudes of these $\ket{l\pm 1}$ states. Because the GW interaction restricts the resulting quantum state to a known parametric form, this information can be extracted via parametrized quantum state tomography. This technique resolves the underlying physical parameters using a reduced set of observables, bypassing the need for exhaustive measurement of the full OAM Hilbert space \cite{PhysRevLett.105.150401,PRXQuantum.6.020346,tonner2026parametricquantumstatetomography}. In this Letter, we adopt natural units ($c=G=\hbar=1$) throughout.

\textit{Detection scheme}---As shown in Refs.~\cite{q38s-k2tq,suppleTomography}, the transition amplitudes of the sideband states $\ket{l\pm 1}$ provide only two independent real DOFs, insufficient to characterize a GW with a single arm. We therefore propose a V-shaped detector (Fig.~\ref{fig.LightPath}), whose two arms provide up to four independent real DOFs, as shown below. Although the scheme applies to arbitrary initial OAM index $l$, we set $l=0$ to simplify optical preparation. A phase modulator prepares the input laser in the zero-OAM Bessel mode $\ket{0}$ \cite{ARLT2000297,PhysRevLett.58.1499,PhysRevA.88.033818,PhysRevA.79.043802}, and a power-recycling cavity enhances its circulating power \cite{Aasi_2015}. A beam splitter (BS) divides the light into paths $\mathrm{P}\in\{\mathrm{A},\mathrm{B}\}$. In each path, a Mach--Zehnder interferometer (MZI) equipped with Dove prisms separates the input and sideband OAM states \cite{leach2002measuring,PhysRevLett.92.013601,PhysRevA.83.053829}; see Sec.~II of the Supplemental Material \cite{suppleTomography} for details. The input state $\ket{0}_{\mathrm{P}}$ enters port $a_1$ and exits exclusively through port $b_2$, then undergoes $N$ round trips in arm $\mathrm{P}$ of length $L$, interacting with the GW.

\begin{figure}[tbhp]
	\centering
	\includegraphics[width=\linewidth]{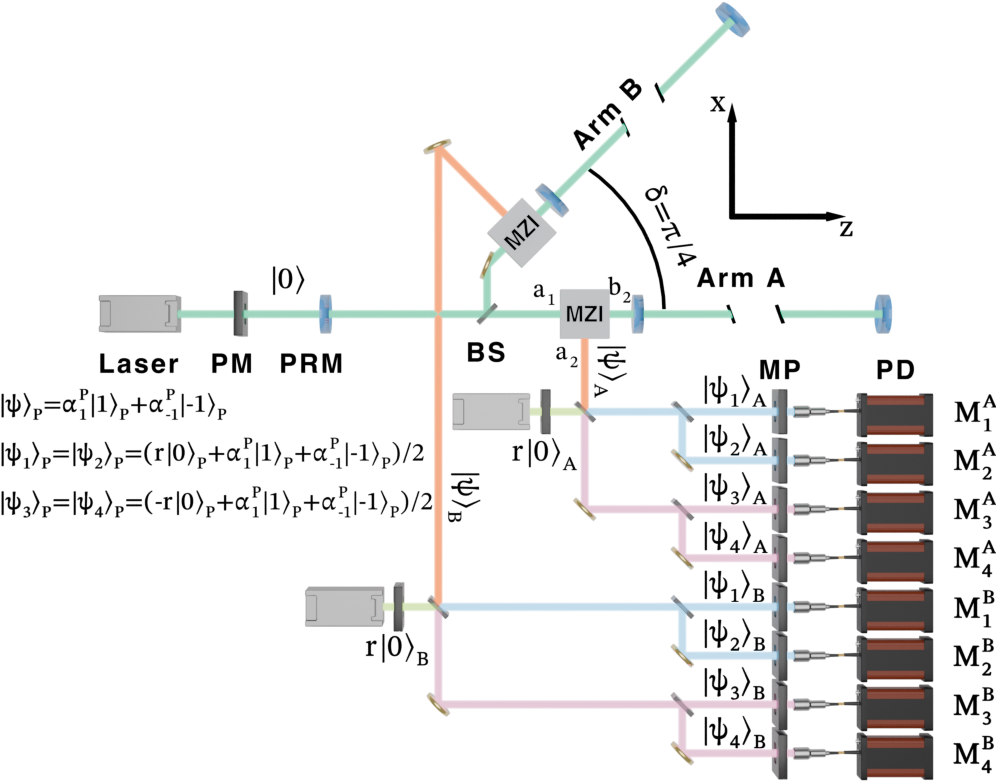}
	\caption{Schematic of the proposed GW detector. Abbreviations: phase modulator (PM), power recycling mirror (PRM), beamsplitter (BS), Mach-Zehnder interferometer (MZI), mode projector (MP), photodetector (PD). Solid black squares indicate breaks in the macroscopic arms. The states selected by the MPs and measured by the corresponding PDs are listed at the lower left.}
	\label{fig.LightPath}
\end{figure}

Arms A and B lie within the $xz$ plane. Arm A is aligned with the $z$ axis, and the angle between the two arms is $\delta$ (Fig.~\ref{fig.LightPath}). Unlike our previous single-arm model \cite{q38s-k2tq}, the two-arm geometry breaks axial symmetry, requiring both polar and azimuthal angles $(\theta,\phi)$ to specify the GW direction. We therefore extend the interaction theory to include $\phi$. To first order in the GW amplitudes, the photon state after interaction is
\begin{equation}
	\ket{0}_{\mathrm{P}}+\alpha^{\mathrm{P}}_{+}\ket{1}_{\mathrm{P}}
	+\alpha^{\mathrm{P}}_{-}\ket{-1}_{\mathrm{P}},\label{eq.AfterInteractionState}
\end{equation}
with transition amplitudes \cite{suppleTomography}
\begin{equation}
	\alpha^{\mathrm{P}}_{\pm}=\sum_{\lambda=+,\times}D^{\mathrm{P}}_{\pm,\lambda}f_{\mathrm{P},\lambda}(t),\qquad \mathrm{P}\in\{\mathrm{A},\mathrm{B}\}.\label{eq.alphaAB}
\end{equation}
The factors $D^{\mathrm{P}}_{\pm,\lambda}$ are linear in $A_\lambda$, with angular dependence determined by combinations of $\theta$, $\phi$, and $\delta$. The GW response function $f_{\mathrm{P},\lambda}(t)$ can be approximately given by\begin{equation}
 -\gamma k_3 NL  \cos\left [k_g L N-\Psi_\lambda+k_g t\right ]\left [1+(f_g/f_p)^2\right ]^{-1/2},
\end{equation}
where $\gamma=k_\perp/k_3$ is the paraxial parameter, $k_\perp$ and $k_3$ are the transverse and longitudinal photon wave numbers, respectively, the exact signal attenuation $\sin(k_g NL)/(k_g NL)$ is approximated by the single-pole transfer function $[1+(f_g/f_p)^2]^{-1/2}$ \cite{9780198570745.001.0001}, $f_g=k_g/(2\pi)$, and $f_p=1/(2\pi NL)$ is the pole frequency \cite{Aasi_2015}. In the low-frequency limit $2Nk_gL\ll 1$, it further simplifies to $ -\gamma k_3 N L \cos(\Psi_\lambda-k_gt)$. Explicit expressions are given in Sec.~I of the Supplemental Material \cite{suppleTomography}. The amplitudes satisfy $\alpha^{\mathrm{P}}_{+}=-(\alpha^{\mathrm{P}}_{-})^*$. For ideal Bessel beams, $\gamma$ is unrestricted within the paraxial regime $\gamma\ll 1$, but these beams require infinite energy. Physically realizable Bessel--Gaussian modes instead have a finite diffraction-free range, $z_{\mathrm{max}}= w_0/\gamma$ \cite{PhysRevLett.58.1499,Liu:26}, $w_0$ is the beam waist. Requiring $z_{\mathrm{max}}= L$ with a realistically large waist $w_0=0.4\,\mathrm{m}$ gives $\gamma \le (0.4\,\mathrm{m})/L$. Unless otherwise stated, we use an optical wavelength $\lambda_{\mathrm{opt}}=1064\,\mathrm{nm}$, $\gamma=0.1$, $A_+=A_\times=10^{-18}$, $\theta=\pi/3$, $\phi=\pi/6$, $\Psi_+=0$, $\Psi_\times=\pi/2$, incident power $P_{\mathrm{BS}}=10\,\mathrm{kW}$ at the BS, $L=4\,\mathrm{m}$, $N=160$, and $\delta=\pi/4$. These parameters give $|\alpha^{\mathrm{P}}_{\pm}|\sim 10^{-10}$.

After exiting arm $\mathrm{P}$, the state in Eq.~\eqref{eq.AfterInteractionState} reenters the MZI through port $b_2$. Photons in $\ket{0}_{\mathrm{P}}$ exit through port $a_1$ and return to the power-recycling cavity, whereas the GW-induced component,
\begin{equation}
	\ket{\psi}_{\mathrm{P}}
	=\alpha^{\mathrm{P}}_{+}\ket{1}_{\mathrm{P}}
	+\alpha^{\mathrm{P}}_{-}\ket{-1}_{\mathrm{P}},
	\label{eq.exitingState}
\end{equation}
exits through port $a_2$ and enters the quantum state tomography stage \cite{suppleTomography}. Since $\alpha^{\mathrm{P}}_{\pm}$ are linear in the GW amplitudes, direct intensity measurements of the $\ket{\pm 1}_{\mathrm{P}}$ modes yield signals quadratic in $A_\lambda$, with $A_\lambda^2\sim 10^{-36}$. The corresponding signal power is approximately $10^{-15}\,\mathrm{W}$ \cite{q38s-k2tq}, or about $10^3$ photons per second. Moreover, $\alpha^{\mathrm{P}}_{+}=-(\alpha^{\mathrm{P}}_{-})^*$ implies equal sideband intensities, so these measurements provide only one independent real DOF per arm and cannot fully characterize the GW signal.

To enhance the detected signal and access both real DOFs, we coherently superpose the sidebands with a stronger reference state $r\ket{0}_{\mathrm{P}}$. In the regime $|\alpha^{\mathrm{P}}_{\pm}|\ll r\ll 1$, the shot-noise-limited sensitivity is independent of $r$, as will be shown in Eq.~\eqref{eq.sensitivity}. Here, we set $r=0.01$. The projection measurements, later defined in Eqs.~(\ref{eq.m1}--\ref{eq.m4}), yield an contribution to the detected power proportional to $rA_\lambda$, with $rA_\lambda\sim 10^{-20}$. An reference laser, phase-locked to the source laser, supplies the reference state with power $r^2P_{\mathrm{BS}}/2=0.5\,\mathrm{W}$. The signal and reference fields are combined at a BS, and each output is split equally by a subsequent BS, yielding the unnormalized mode states \cite{suppleTomography}
\begin{align}
	\ket{\psi_1}_{\mathrm{P}}=\ket{\psi_2}_{\mathrm{P}}
	&=\frac{1}{2}\left(
	r\ket{0}_{\mathrm{P}}
	+\alpha^{\mathrm{P}}_{+}\ket{1}_{\mathrm{P}}
	+\alpha^{\mathrm{P}}_{-}\ket{-1}_{\mathrm{P}}
	\right),
	\label{eq.psi1}\\
	\ket{\psi_3}_{\mathrm{P}}=\ket{\psi_4}_{\mathrm{P}}
	&=\frac{1}{2}\left(
	-r\ket{0}_{\mathrm{P}}
	+\alpha^{\mathrm{P}}_{+}\ket{1}_{\mathrm{P}}
	+\alpha^{\mathrm{P}}_{-}\ket{-1}_{\mathrm{P}}
	\right).
	\label{eq.psi3}
\end{align}
Each $\ket{\psi_m}_{\mathrm{P}}$, with $m=1,\dots,4$, is projected by the corresponding mode projector and measured by photodetector $M^{\mathrm{P}}_m$.

% \section{Parametrized quantum state tomography for extracting GW wave parameters}
\textit{Parametrized quantum tomography for characterizing a GW signal}---We reconstruct $\ket{\psi}_{\mathrm{P}}$ using eight projection measurements, four per arm. The projection states are $\ket{\Phi_1}_{\mathrm{P}}=(\ket{0}_{\mathrm{P}}+\ket{1}_{\mathrm{P}})/\sqrt{2}$, $\ket{\Phi_2}_{\mathrm{P}}=(\ket{0}_{\mathrm{P}}+i\ket{1}_{\mathrm{P}})/\sqrt{2}$, $\ket{\Phi_3}_{\mathrm{P}}=(-\ket{0}_{\mathrm{P}}+\ket{-1}_{\mathrm{P}})/\sqrt{2}$, and $\ket{\Phi_4}_{\mathrm{P}}=(-\ket{0}_{\mathrm{P}}-i\ket{-1}_{\mathrm{P}})/\sqrt{2}$. Each $\ket{\psi_m}_{\mathrm{P}}$ passes through a mode projector, such as a spatial light modulator, that selects $\ket{\Phi_m}_{\mathrm{P}}$. The measured power is proportional to $|\bra{\Phi_m}_{\mathrm{P}}\ket{\psi_m}_{\mathrm{P}}|^2$. Neglecting terms quadratic in $\alpha^{\mathrm{P}}_{\pm}$, the measured powers are \cite{suppleTomography}
\begin{align}
	M^{\mathrm{P}}_1
	&\approx P_{\mathrm{BS}}\left[r^2+2r\operatorname{Re}(\alpha^{\mathrm{P}}_+)\right]/16,\label{eq.m1}\\
	M^{\mathrm{P}}_2
	&\approx P_{\mathrm{BS}}\left[r^2+2r\operatorname{Im}(\alpha^{\mathrm{P}}_+)\right]/16,\label{eq.m2}\\
	M^{\mathrm{P}}_3
	&\approx P_{\mathrm{BS}}\left[r^2+2r\operatorname{Re}(\alpha^{\mathrm{P}}_-)\right]/16,\label{eq.m3}\\
	M^{\mathrm{P}}_4
	&\approx P_{\mathrm{BS}}\left[r^2-2r\operatorname{Im}(\alpha^{\mathrm{P}}_-)\right]/16.\label{eq.m4}
\end{align}
These measurements determine the complex transition amplitudes $\alpha^{\mathrm{P}}_\pm$ and thus reconstruct $\ket{\psi}_{\mathrm{P}}$ in both arms.

The eight outputs contain at most four independent real DOFs. The orthogonal $+$ and $\times$ polarization tensors are related by a spatial rotation of $\pi/4$ about the GW propagation direction, reflecting the spin-2 nature of the field \cite{misner1973gravitation}. Measurement degeneracies depend on how these tensors couple to the two-arm geometry. For the response in Eq.~\eqref{eq.alphaAB}, the number of independent real DOFs reduces to two for parallel or antiparallel arms ($\delta=0$ or $\pi$) and three for perpendicular arms ($\delta=\pi/2$). Four independent real DOFs are generically accessible when the arms are neither parallel nor perpendicular. We henceforth choose a V-shaped configuration with $\delta=\pi/4$, avoiding both degenerate arm geometries.

In this configuration, the eight outputs form four balanced channels that retain the four independent real DOFs while suppressing common-mode noise \cite{Painchaud:09}. Using $\alpha^{\mathrm{P}}_- = -(\alpha^{\mathrm{P}}_+)^*$, we obtain
\begin{align}
	R_{\mathrm{P}}
	&=M^{\mathrm{P}}_1-M^{\mathrm{P}}_3
	\approx P_{\mathrm{BS}}r
	\operatorname{Re}(\alpha^{\mathrm{P}}_+)/4,\\
	I_{\mathrm{P}}
	&=M^{\mathrm{P}}_2-M^{\mathrm{P}}_4
	\approx P_{\mathrm{BS}}r
	\operatorname{Im}(\alpha^{\mathrm{P}}_+)/4.
\end{align}
Under ideal balancing, these differences cancel the reference DC contribution and its common-mode fluctuations while retaining the two GW-induced signal quadratures in each arm.

The reconstructed states provide access to the wave parameters. We assume that the signal duration $t_d$ greatly exceeds the optical storage time, $t_d\gg 2NL$. We define $O_1=\operatorname{Re}(\alpha^{\mathrm{A}}_+)$, $O_2=\operatorname{Im}(\alpha^{\mathrm{A}}_+)$, $O_3=\operatorname{Re}(\alpha^{\mathrm{B}}_+)$, and $O_4=\operatorname{Im}(\alpha^{\mathrm{B}}_+)$, and introduce the intermediate variables $Y = \sqrt{2}O_4 - O_2$, $c_0 = Y^2 O_1 - 2Y O_3 O_2 - O_1 O_2^2$, $c_1 = -O_2(Y^2 + 4O_3^2 + 4O_1^2 + O_2^2)$, $c_2 = (2Y^2 - 4O_3^2)O_1 - 6Y O_3 O_2 - 4O_1(O_1^2 + O_2^2)$, and $c_3 = -2(Y^2 O_2 + 2Y O_3 O_1 + 2O_1^2 O_2)$. Then, the azimuthal angle $\phi$ can be found by solving the equation
\begin{equation}
	c_3\tan^3\phi+c_2\tan^2\phi+c_1\tan\phi+c_0=0.	\label{eq.SolvePhi}
\end{equation}
The polar angle $\theta$ satisfies
\begin{equation}
	\tan\theta=\frac{(2O_3O_2-YO_1)\cos\phi+(YO_2+2O_3O_1)\sin\phi}{
		(O_1\cos\phi+O_2\sin\phi)(2O_1\sin\phi+O_2\cos\phi)}.
\end{equation}
The two effective amplitudes are
\begin{align}
	\tilde{A}_\times &= \frac{O_1 \sin\phi - O_2 \cos\phi}{\gamma k_3 N L \sin\theta},\\
	\tilde{A}_+ &= -\frac{O_1 \cos\phi + O_2 \sin\phi}{\gamma k_3 N L \sin\theta \cos\theta}.\label{eq.SolveAp}
\end{align}
Frequency-domain analysis of $\tilde{A}_\lambda(t)$ can then determine the frequency $k_g$, amplitudes $A_\lambda$, and phases $\Psi_\lambda$. Recovering both polarization waveforms and the propagation direction provides additional observables for source-parameter estimation.

\begin{figure}[tbhp]
	\centering
	\includegraphics[width=\linewidth]{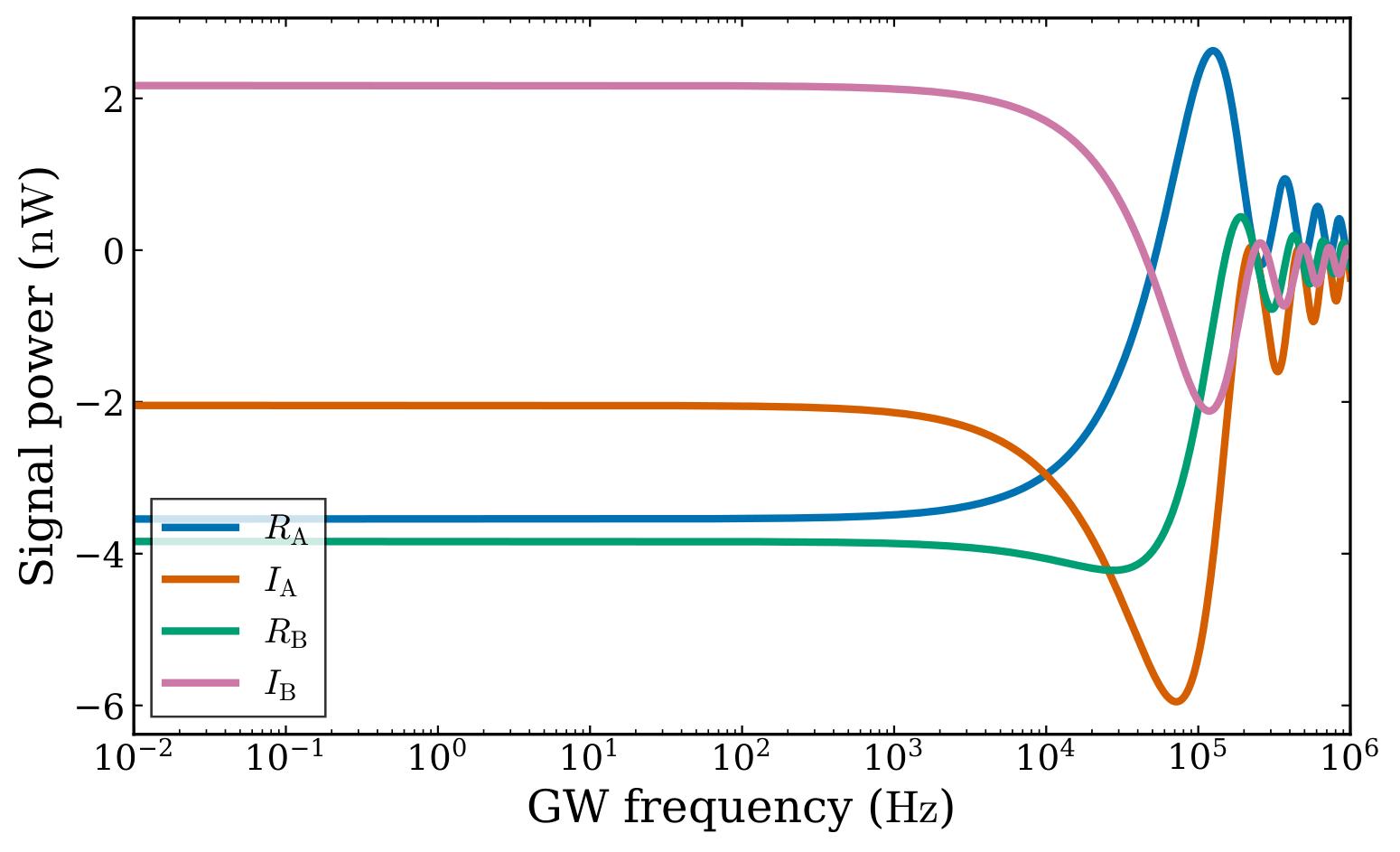}
	\caption{Balanced signal powers $R_{\mathrm{P}}$ and $I_{\mathrm{P}}$ at $t=0$ as functions of GW frequency. Their magnitudes reach the nanowatt level at low frequencies. The incident power is $P_{\mathrm{BS}}=10\,\mathrm{kW}$; the remaining parameters are specified in the main text.}
	\label{fig.SignalPower}
\end{figure}

%\section{Shot noise budget and antenna patterns}
\textit{Shot-noise budget and antenna patterns}---Each photodetector measures a DC reference contribution ($P_{\mathrm{BS}}r^2/16$) and a GW-induced signal ($\pm R_{\mathrm{P}}/2$ or $\pm I_{\mathrm{P}}/2$). For the chosen $P_{\mathrm{BS}}$ and $r$, the DC power is $62.5\,\mathrm{mW}$ per photodetector. The GW-induced signal is superimposed on this DC power during DC readout \cite{fricke2012dc,PhysRevLett.126.221301,PhysRevD.93.112004}. Figure~\ref{fig.SignalPower} shows $R_{\mathrm{P}}$ and $I_{\mathrm{P}}$ at $t=0$. At low frequencies, their characteristic magnitudes scale as $P_{\mathrm{BS}}r|\alpha^{\mathrm{P}}_{\pm}|\propto P_{\mathrm{BS}}rA_\lambda NL\gamma k_3$ and reach the nanowatt level. At frequencies above the pole frequency $f_p=7.46\times 10^4$, the balanced signal powers exhibit oscillatory behavior with an overall decreasing envelope.

Our scheme detects GW-induced OAM transitions rather than differential arm-length changes. In the low-frequency regime, $\alpha^{\mathrm{P}}_{\pm}\propto L$, so a small displacement $\Delta L$ changes the linear signal-power contribution by a fraction $\Delta L/L$. For $L=4\,\mathrm{m}$ and a representative signal power of $1\,\mathrm{nW}$, as shown in Fig.~\ref{fig.SignalPower}, a displacement of $\Delta L=1\,\mu\mathrm{m}$ produces a power fluctuation of approximately $2.5\times10^{-16}\,\mathrm{W}$. This estimation indicates weak coupling of displacement noise through variations in the interaction length, including displacements driven by radiation pressure, coating Brownian noise, suspension thermal noise, seismic motion, and gravity gradients \cite{PhysRevD.93.112004,PhysRevLett.116.131103,PhysRevD.103.122004}. Below, we evaluate the shot-noise-limited sensitivity. Other noise sources are briefly discussed in the conclusion.

For each photodetector, the mean photon count during an observation time $T$ is $\bar N_\gamma=\bar P_{\mathrm{obs}}T/\omega$, where $\bar P_{\mathrm{obs}}$ is the time-averaged measured power and $\omega$ is the photon frequency. Due to Poisson statistics, the photon number variance is $\sigma^2=\bar N_\gamma$. Assuming independent measurements across the eight photodetectors, the total variance is $\sigma^2_{\text{total}}=8\sigma^2$, corresponding to a shot-noise power fluctuation $P_{\text{shot}}=\sigma_{\text{total}} \omega/T$. Without a GW signal, the DC carrier power per photodetector is $\bar P_{\text{obs}}=P_{\text{BS}} r^2/16 $. This yields $P_{\text{shot}}=[r^2 P_{\text{BS}}\omega/(2T)]^{1/2}$. Because the photodetectors measure either the real or imaginary parts of $\alpha^{\text{P}}_{\pm }$, and $|\operatorname{Re}(\alpha^{\text{P}}_{\pm })|+|\operatorname{Im}(\alpha^{\text{P}}_{\pm })|\ge |\alpha^{\text{P}}_{\pm}| $, the lower bound for the GW-induced total power fluctuation is $P_{\text{GW}}=4(P_{\text{BS}}/2)(r/4)|\alpha^{\text{P}}_{\pm }|=r|\alpha^{\text{P}}_{\pm }|P_{\text{BS}}/2$. The signal-to-noise ratio ($S/N$) is thus \cite{9780198570745.001.0001,moore2014gravitational}
\begin{equation}
	\frac{S}{N}=\frac {P_{\text{GW}}}{P_{\text{shot}}}=\frac {\gamma N L h_0 (T\omega P_{\text{BS}}/2)^{1/2}}{[1+(f_{g}/f_p)^2]^{1/2}},
\end{equation}
where $h_0=|(\tilde A_\times +i\tilde A_+\cos\theta)\sin\theta|$, and the $S/N$ is evaluated at $\theta=\pi/2$. The shot-noise-limited strain sensitivity $S^{1/2}_n(f_g)=h_0 T^{1/2}/(S/N)$ is therefore \cite{9780198570745.001.0001,moore2014gravitational}
\begin{align}
	&S^{1/2}_n(f_g)=\frac {1}{\gamma N L}\left (\frac {\lambda_{\mathrm{opt}}}{\pi P_{\text{BS}}} \right )^{1/2} \left [1+(f_g/f_p)^2\right ]^{1/2}\nonumber \\ 
	\approx & 1.6\times 10^{-20} \left(\frac {0.1}{\gamma}\right) \left(\frac {160}{N}\right) \left(\frac {4\text{ m}}{L}\right) \left(\frac {10\text{ kW}}{P_{\text{BS}}} \right )^{1/2}\nonumber \\
	&\times \left [1+(f_g/f_p)^2\right ]^{1/2}.\nonumber \\
	= & 1.6\times 10^{-20}  \left(\frac {160}{N}\right) \left(\frac {0.4\text{ m}}{w_0}\right) \left(\frac {10\text{ kW}}{P_{\text{BS}}} \right )^{1/2}\nonumber \\
	&\times \left [1+(2\pi f_g  NL/c)^2\right ]^{1/2}.	\label{eq.sensitivity}
\end{align}

\begin{figure}[tbhp]
	\centering
	\includegraphics[width=\linewidth]{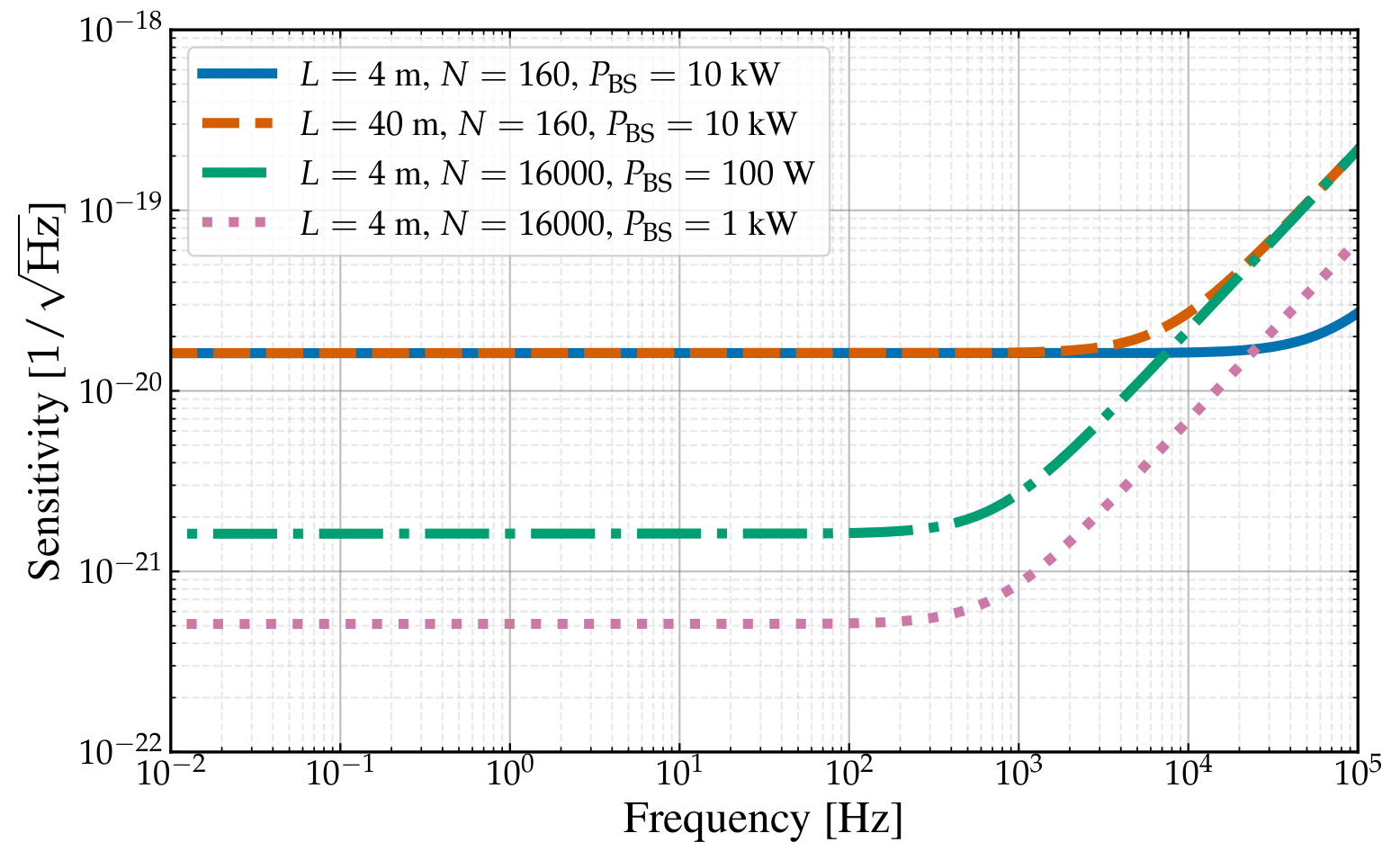}
	 \caption{Shot-noise-limited strain sensitivities of the proposed detector for four configurations. Blue solid: $L=4\,\mathrm{m}$, $N=160$, $P_{\mathrm{BS}}=10\,\mathrm{kW}$, and $\gamma=0.1$. Orange dashed: $L=40\,\mathrm{m}$, $N=160$, $P_{\mathrm{BS}}=10\,\mathrm{kW}$, and $\gamma=0.01$. Green dash-dotted: $L=4\,\mathrm{m}$, $N=16000$, $P_{\mathrm{BS}}=100\,\mathrm{W}$, and $\gamma=0.1$. Purple dotted: $L=4\,\mathrm{m}$, $N=16000$, $P_{\mathrm{BS}}=1\,\mathrm{kW}$, and $\gamma=0.1$. Increasing the effective interaction time improves the low-frequency sensitivity at the expense of bandwidth, while increasing the incident power further reduces the shot-noise floor. The purple curve represents an idealized high-power benchmark. Other parameters are specified in the main text.}
	\label{fig.Sensitivity}
\end{figure}

For $L=4\,\mathrm{m}$, $N=160$, $P_{\mathrm{BS}}=10\,\mathrm{kW}$, and $\gamma=0.1$, the shot-noise-limited strain sensitivity reaches $1.6\times10^{-20}\,\mathrm{Hz}^{-1/2}$ at low frequencies, with a pole frequency of $f_p=7.46\times10^4\,\mathrm{Hz}$ (blue curve in Fig.~\ref{fig.Sensitivity}). Increasing $L$ to $40\,\mathrm{m}$ requires a smaller $\gamma=0.01$ to accommodate the more restrictive diffraction-range constraint. Since $\gamma L$ remains unchanged, the sensitivity floor is preserved, while $f_p$ decreases to $7.46\times10^3\,\mathrm{Hz}$ (orange curve). Alternatively, increasing $N$ to $16000$ while reducing $P_{\mathrm{BS}}$ to $100\,\mathrm{W}$ improves the sensitivity floor to $1.6\times10^{-21}\,\mathrm{Hz}^{-1/2}$, with $f_p=7.46\times10^2\,\mathrm{Hz}$ (green curve). This configuration maintains the same circulating power of $0.8\,\mathrm{MW}$ as the first configuration. Further increasing $P_{\mathrm{BS}}$ to $1\,\mathrm{kW}$ lowers the sensitivity floor to $5.1\times10^{-22}\,\mathrm{Hz}^{-1/2}$ (purple curve). These results demonstrate that increasing the effective interaction time improves low-frequency sensitivity at the expense of bandwidth. All four configurations exhibit nearly constant shot-noise-limited sensitivity over the $0.1$--$10\,\mathrm{Hz}$ band, highlighting the potential of the proposed scheme for mid-frequency GW detection.

The last configuration with $P_{\mathrm{BS}}=1\,\mathrm{kW}$ corresponds to a circulating power of approximately $8\,\mathrm{MW}$, exceeding the nominal design values of existing and proposed GW interferometers. We emphasize that this high-power configuration serves as an optimistic theoretical benchmark rather than an immediately realizable experimental design. Nevertheless, the configuration with $P_{\mathrm{BS}}=100\,\mathrm{W}$ already achieves a sensitivity floor of $1.6\times10^{-21}\,\mathrm{Hz}^{-1/2}$ at a circulating power comparable to the Advanced LIGO design value. Thus, the sensitivity enhancement does not rely exclusively on the high-power configuration, while a detailed assessment of experimental feasibility remains for future investigation.

\begin{figure*}[tbhp]
	\centering
	\includegraphics[width=0.23\linewidth]{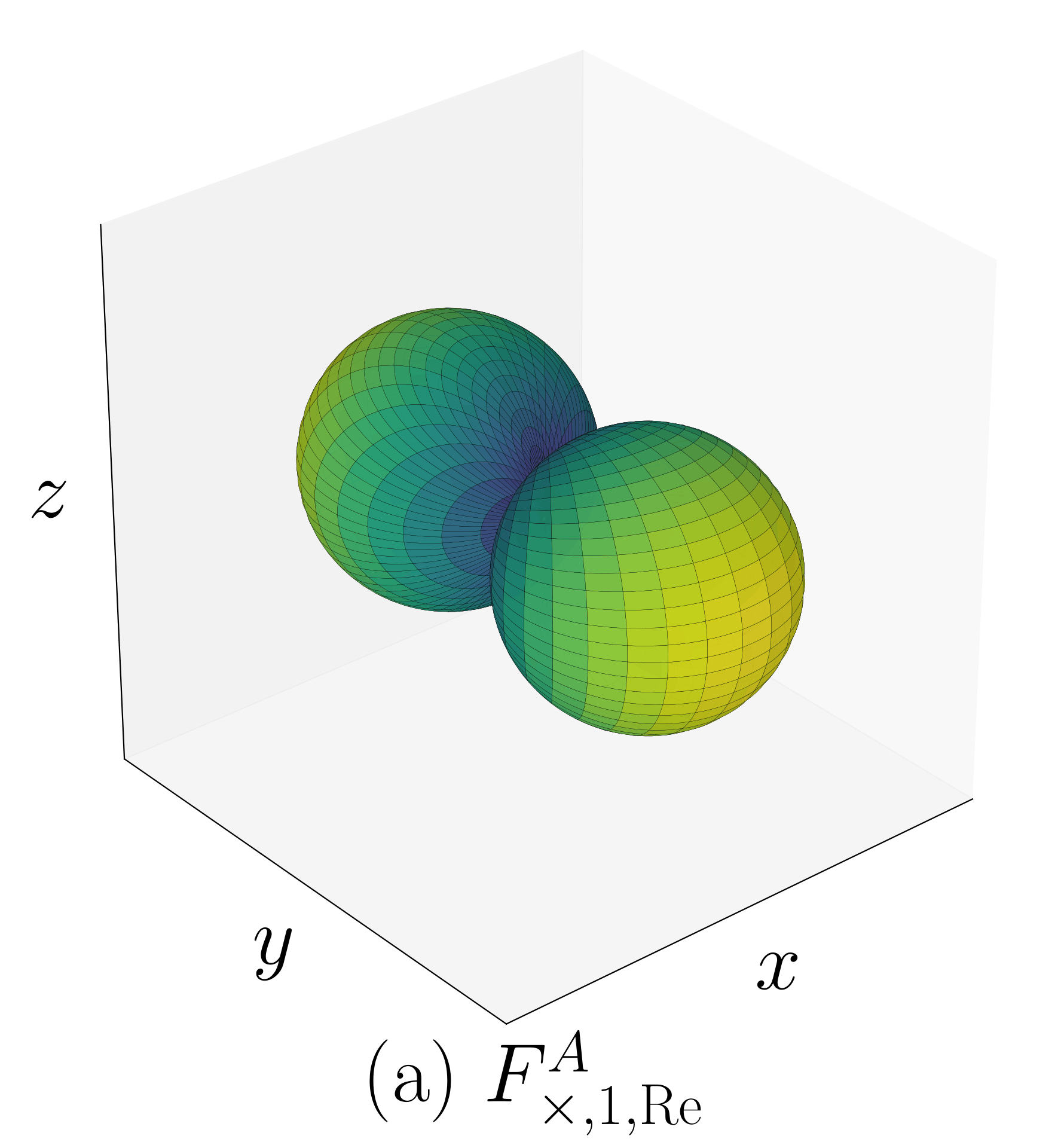}\hfill
	\includegraphics[width=0.23\linewidth]{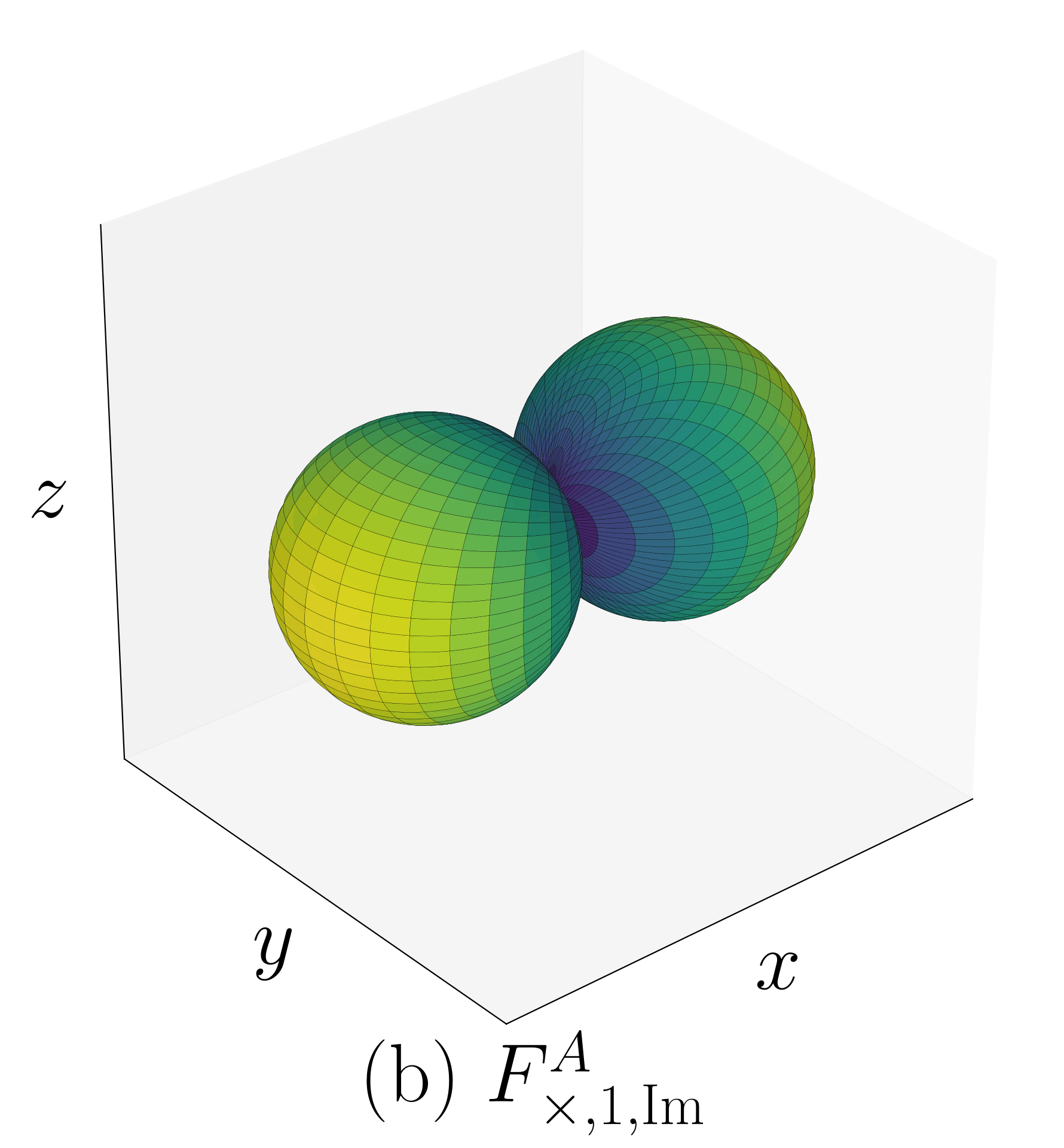}\hfill
	\includegraphics[width=0.23\linewidth]{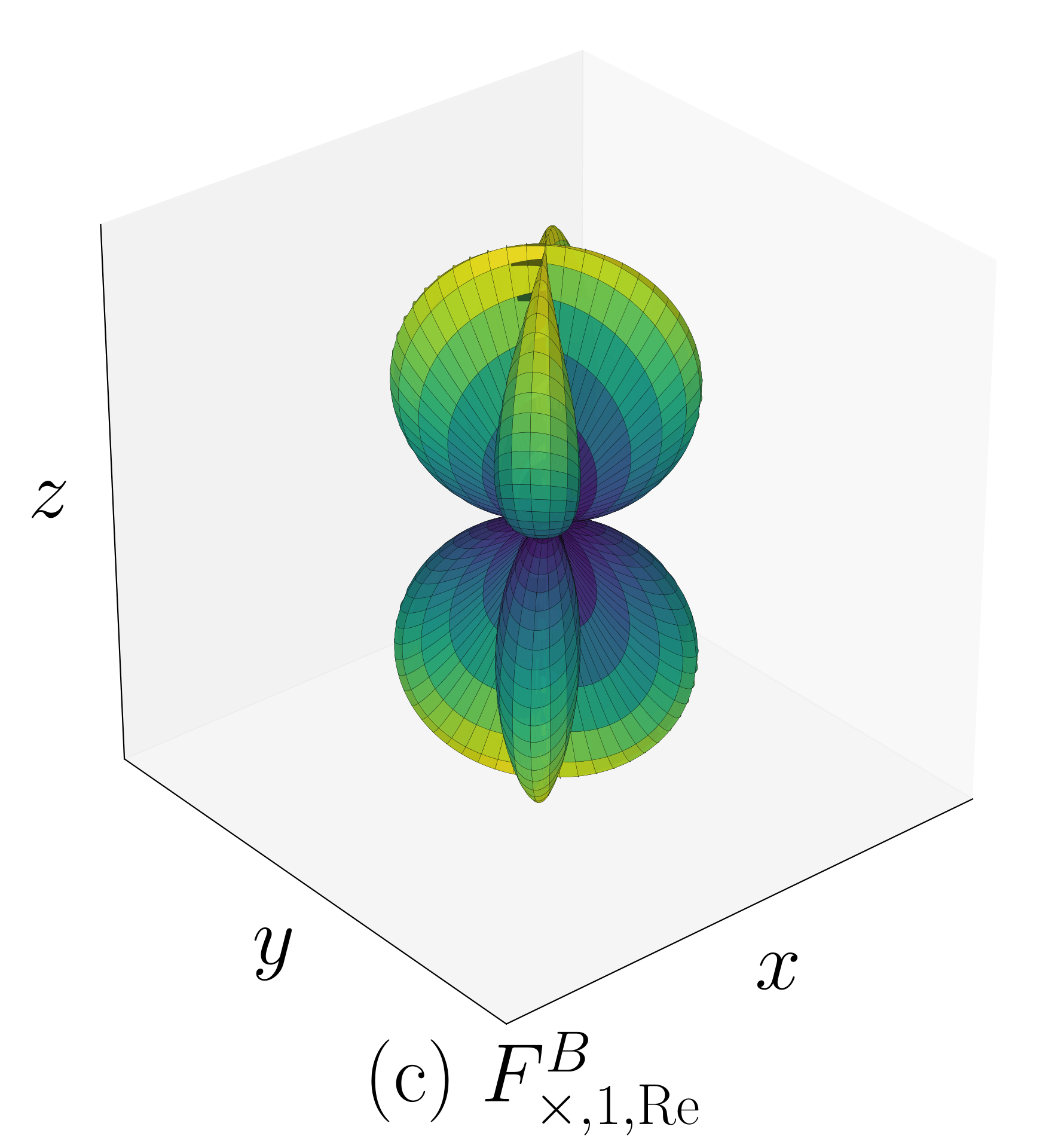}\hfill
	\includegraphics[width=0.23\linewidth]{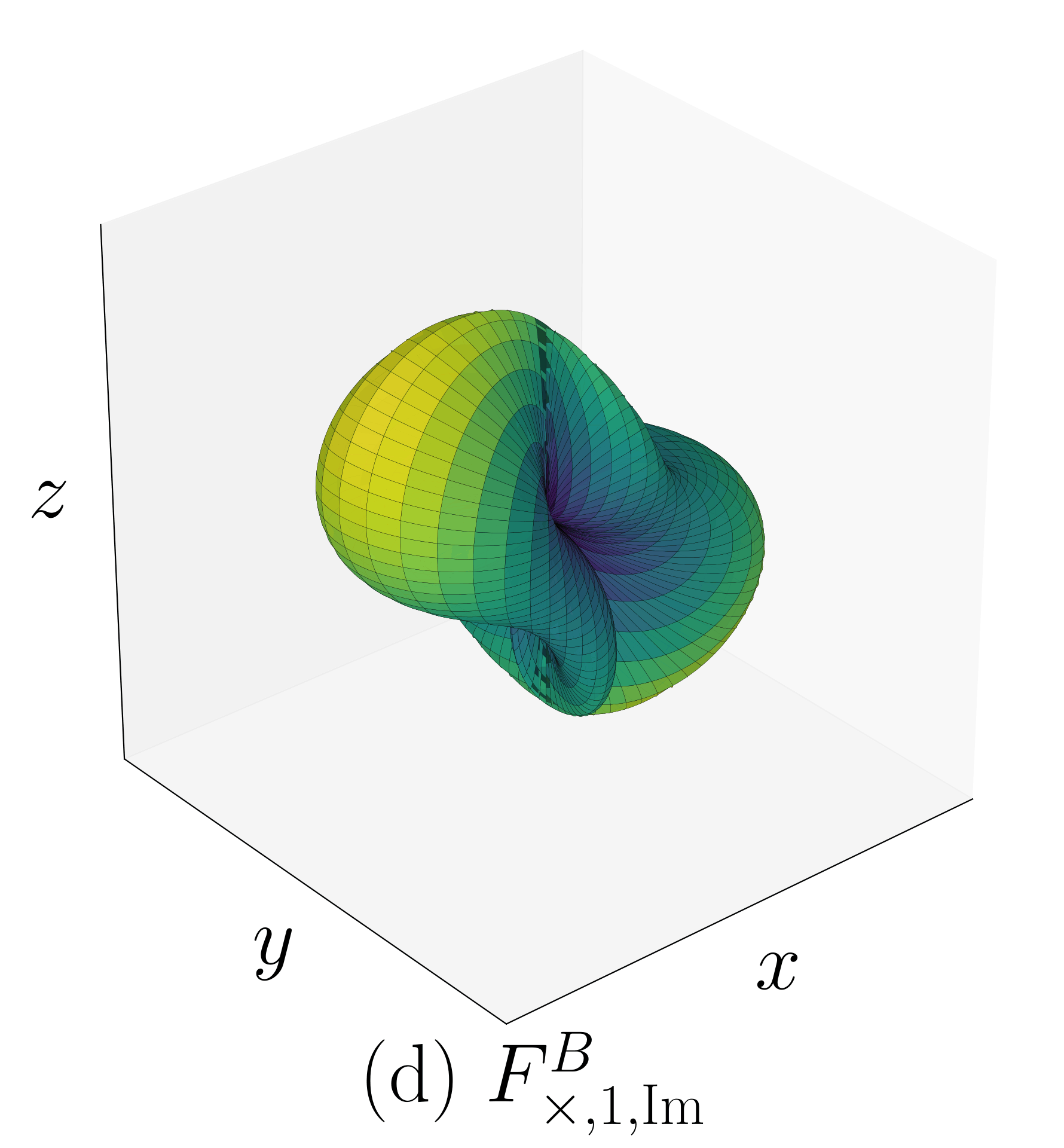}
	\caption{Antenna patterns for the $\times$-polarized GW. (a) $|F^{\text{A}}_{\times,1,\operatorname{Re}}|$ exhibits spatial nulls at $\theta=0$ as well as $\phi=0$ and $\pi$. The azimuthal nulls ($\phi$) are covered by the complementary geometry of (b) $|F^{\text{A}}_{\times,1,\operatorname{Im}}|$, while the polar nulls ($\theta$) are covered by (c) $|F^{\text{B}}_{\times,1,\operatorname{Re}}|$. The pattern for (d) $|F^{\text{B}}_{\times,1,\operatorname{Im}}|$ provides additional complex angular dependence. Together, these patterns ensure comprehensive angular sensitivity.}\label{fig.AntennaPattern}
\end{figure*}

In Michelson interferometers, the measured scalar strain $h(t)=F_+h_+(t)+F_\times h_\times(t)$ is a linear projection of the two GW polarizations ($h_+$ and $h_\times$) weighted by the detector antenna patterns ($F_+$ and $F_\times$) \cite{abbott2016observation,PhysRevD.58.063001,PhysRevLett.116.241102,Abbott_2017}. In our scheme, the wave parameters are reconstructed using four independent DOFs of the transition amplitudes. Factoring in the two GW polarizations yields 8 effective antenna patterns, all of which can be deduced from the real or imaginary parts of the transition amplitudes $\alpha^{\text{P}}_{\pm }$. In the low-frequency limit ($2Nk_gL\ll 1$), $\operatorname{Re}(\alpha^{\text{P}}_{\pm })=C(\tilde A_+ F^{\text{P}}_{+,\pm 1,\operatorname{Re}}+ \tilde A_\times  F^{\text{P}}_{\times,\pm 1,\operatorname{Re}})$ and $\operatorname{Im}(\alpha^{\text{P}}_{\pm })=C(\tilde A_+ F^{\text{P}}_{+,\pm 1,\operatorname{Im}}+ \tilde A_\times  F^{\text{P}}_{\times,\pm 1,\operatorname{Im}})$, where $C=  \gamma k_3 NL$ is a normalization factor and we denote these patterns as $F^{\text{P}}_{\sigma,l,M}$ with $\sigma=+,\times$, $l=\pm 1$, and $M=\operatorname{Re}, \operatorname{Im}$. For example, isolating the $\times$-polarization (with $A_+=0$) yields $|F^{\text{A}}_{\times,1,\operatorname{Re}}|=\sin\theta\sin\phi$, $|F^{\text{A}}_{\times,1,\operatorname{Im}}|=\sin\theta\cos\phi$, $|F^{\text{B}}_{\times,1,\operatorname{Re}}|=\cos\theta\cos\phi\sin\phi$, and $|F^{\text{B}}_{\times,1,\operatorname{Im}}|=(\sin\theta\cos\phi-\cos\theta\cos2\phi)/\sqrt 2$ (Fig. \ref{fig.AntennaPattern}). Although individual patterns exhibit spatial nulls, their complementary geometries provide comprehensive sky coverage, preventing signal loss for arbitrary GW propagation direction.

% \section{Discussion}
\textit{Discussion}---We exploit GW-induced quantum transitions between photon OAM states as a novel detection method for GWs. By reconstructing the four complex transition amplitudes via parametrized quantum tomography, we extract the wave parameters using Eqs. (\ref{eq.SolvePhi}--\ref{eq.SolveAp}). Thus, we demonstrate that a single V-shaped detector can fully resolve the spatial and polarization information of a GW within the OAM Hilbert space.

Our detection scheme offers several advantages. First, it does not rely on measuring differential arm lengths, rendering it robust against displacement noise. We establish its theoretical shot-noise-limited sensitivity while accounting for the finite diffraction-free range of realizable Bessel--Gaussian modes, which constrains $\gamma L$. For $\gamma=0.1$, $L=4\,\mathrm{m}$, $N=160$, the projected low-frequency sensitivity is $1.6\times10^{-20}\,\mathrm{Hz}^{-1/2}$. Increasing $N$ to $16000$ while reducing $P_{\mathrm{BS}}$ to $1\,\mathrm{kW}$ improves the sensitivity floor to $5.1\times10^{-22}\,\mathrm{Hz}^{-1/2}$ while reducing the bandwidth. Second, the resulting sensitivity curves demonstrate that our scheme can probe a broader frequency band, extending particularly into the mid-frequency gap (0.1 to 10 Hz). Third, the complementary spatial geometries of our antenna patterns ensure comprehensive sky coverage, eliminating geometric blind spots.

Although our method is free from displacement noise, its sensitivity remains susceptible to other noise sources, including laser-intensity noise, which affects $P_{\mathrm{BS}}$ and $r$; laser-phase noise, which perturbs the phase of $r$; and mode-impurity noise, which contaminates the transition amplitudes $\alpha_\pm^{\mathrm{P}}$. We find that, with balanced detection, the proposed detector can suppress the technical noises sufficiently to bring the sensitivity close to the shot-noise limit, yielding a maximum SNR of approximately $1.4\times10^6$ for GW signals evaluated over the frequency band from $1~\mathrm{mHz}$ to $100~\mathrm{Hz}$. A detailed noise-budget analysis, together with a systematic evaluation of the corresponding SNR and detectability, will be reported elsewhere.

\section*{Acknowledgments}
This work is supported by National Natural Science Foundation of China (125B2103).

\section*{Data availability}
The data that support the findings of this article are not publicly available. The data are available from the authors upon reasonable request.


\begin{thebibliography}{39}%
	\makeatletter
	\providecommand \@ifxundefined [1]{%
		\@ifx{#1\undefined}
	}%
	\providecommand \@ifnum [1]{%
		\ifnum #1\expandafter \@firstoftwo
		\else \expandafter \@secondoftwo
		\fi
	}%
	\providecommand \@ifx [1]{%
		\ifx #1\expandafter \@firstoftwo
		\else \expandafter \@secondoftwo
		\fi
	}%
	\providecommand \natexlab [1]{#1}%
	\providecommand \enquote  [1]{``#1''}%
	\providecommand \bibnamefont  [1]{#1}%
	\providecommand \bibfnamefont [1]{#1}%
	\providecommand \citenamefont [1]{#1}%
	\providecommand \href@noop [0]{\@secondoftwo}%
	\providecommand \href [0]{\begingroup \@sanitize@url \@href}%
	\providecommand \@href[1]{\@@startlink{#1}\@@href}%
	\providecommand \@@href[1]{\endgroup#1\@@endlink}%
	\providecommand \@sanitize@url [0]{\catcode `\\12\catcode `\$12\catcode
		`\&12\catcode `\#12\catcode `\^12\catcode `\_12\catcode `\%12\relax}%
	\providecommand \@@startlink[1]{}%
	\providecommand \@@endlink[0]{}%
	\providecommand \url  [0]{\begingroup\@sanitize@url \@url }%
	\providecommand \@url [1]{\endgroup\@href {#1}{\urlprefix }}%
	\providecommand \urlprefix  [0]{URL }%
	\providecommand \Eprint [0]{\href }%
	\providecommand \doibase [0]{https://doi.org/}%
	\providecommand \selectlanguage [0]{\@gobble}%
	\providecommand \bibinfo  [0]{\@secondoftwo}%
	\providecommand \bibfield  [0]{\@secondoftwo}%
	\providecommand \translation [1]{[#1]}%
	\providecommand \BibitemOpen [0]{}%
	\providecommand \bibitemStop [0]{}%
	\providecommand \bibitemNoStop [0]{.\EOS\space}%
	\providecommand \EOS [0]{\spacefactor3000\relax}%
	\providecommand \BibitemShut  [1]{\csname bibitem#1\endcsname}%
	\let\auto@bib@innerbib\@empty
	%</preamble>
	\bibitem [{\citenamefont {Abbott}\ \emph
		{et~al.}(2016{\natexlab{a}})\citenamefont {Abbott}, \citenamefont {Abbott},
		\citenamefont {Abbott}, \citenamefont {Abernathy}, \citenamefont {Acernese},
		\citenamefont {Ackley}, \citenamefont {Adams}, \citenamefont {Adams},
		\citenamefont {Addesso}, \citenamefont {Adhikari} \emph
		{et~al.}}]{abbott2016observation}%
	\BibitemOpen
	\bibfield  {author} {\bibinfo {author} {\bibfnamefont {B.~P.}\ \bibnamefont
			{Abbott}}, \bibinfo {author} {\bibfnamefont {R.}~\bibnamefont {Abbott}},
		\bibinfo {author} {\bibfnamefont {T.~D.}\ \bibnamefont {Abbott}}, \bibinfo
		{author} {\bibfnamefont {M.~R.}\ \bibnamefont {Abernathy}}, \bibinfo {author}
		{\bibfnamefont {F.}~\bibnamefont {Acernese}}, \bibinfo {author}
		{\bibfnamefont {K.}~\bibnamefont {Ackley}}, \bibinfo {author} {\bibfnamefont
			{C.}~\bibnamefont {Adams}}, \bibinfo {author} {\bibfnamefont
			{T.}~\bibnamefont {Adams}}, \bibinfo {author} {\bibfnamefont
			{P.}~\bibnamefont {Addesso}}, \bibinfo {author} {\bibfnamefont {R.~X.}\
			\bibnamefont {Adhikari}}, \emph {et~al.},\ }\bibfield  {title} {\bibinfo
		{title} {Observation of gravitational waves from a binary black hole
			merger},\ }\href {https://doi.org/10.1103/PhysRevLett.116.061102} {\bibfield
		{journal} {\bibinfo  {journal} {Phys. Rev. Lett.}\ }\textbf {\bibinfo
			{volume} {116}},\ \bibinfo {pages} {061102} (\bibinfo {year}
		{2016}{\natexlab{a}})}\BibitemShut {NoStop}%
	\bibitem [{\citenamefont {Abbott}\ \emph
		{et~al.}(2017{\natexlab{a}})\citenamefont {Abbott}, \citenamefont {Abbott},
		\citenamefont {Abbott} \emph {et~al.}}]{PhysRevLett.119.161101}%
	\BibitemOpen
	\bibfield  {author} {\bibinfo {author} {\bibfnamefont {B.~P.}\ \bibnamefont
			{Abbott}}, \bibinfo {author} {\bibfnamefont {R.}~\bibnamefont {Abbott}},
		\bibinfo {author} {\bibfnamefont {T.~D.}\ \bibnamefont {Abbott}}, \emph
		{et~al.} (\bibinfo {collaboration} {LIGO Scientific Collaboration and Virgo
			Collaboration}),\ }\bibfield  {title} {\bibinfo {title} {Gw170817:
			Observation of gravitational waves from a binary neutron star inspiral},\
	}\href {https://doi.org/10.1103/PhysRevLett.119.161101} {\bibfield  {journal}
		{\bibinfo  {journal} {Phys. Rev. Lett.}\ }\textbf {\bibinfo {volume} {119}},\
		\bibinfo {pages} {161101} (\bibinfo {year} {2017}{\natexlab{a}})}\BibitemShut
	{NoStop}%
	\bibitem [{\citenamefont {Abbott}\ \emph {et~al.}(2023)\citenamefont {Abbott},
		\citenamefont {Abbott}, \citenamefont {Acernese} \emph
		{et~al.}}]{PhysRevX.13.041039}%
	\BibitemOpen
	\bibfield  {author} {\bibinfo {author} {\bibfnamefont {R.}~\bibnamefont
			{Abbott}}, \bibinfo {author} {\bibfnamefont {T.~D.}\ \bibnamefont {Abbott}},
		\bibinfo {author} {\bibfnamefont {F.}~\bibnamefont {Acernese}}, \emph
		{et~al.} (\bibinfo {collaboration} {LIGO Scientific Collaboration, Virgo
			Collaboration, and KAGRA Collaboration}),\ }\bibfield  {title} {\bibinfo
		{title} {Gwtc-3: Compact binary coalescences observed by ligo and virgo
			during the second part of the third observing run},\ }\href
	{https://doi.org/10.1103/PhysRevX.13.041039} {\bibfield  {journal} {\bibinfo
			{journal} {Phys. Rev. X}\ }\textbf {\bibinfo {volume} {13}},\ \bibinfo
		{pages} {041039} (\bibinfo {year} {2023})}\BibitemShut {NoStop}%
	\bibitem [{\citenamefont {Maggiore}(2008)}]{maggiore2008gravitational}%
	\BibitemOpen
	\bibfield  {author} {\bibinfo {author} {\bibfnamefont {M.}~\bibnamefont
			{Maggiore}},\ }\href@noop {} {\emph {\bibinfo {title} {Gravitational
				waves}}},\ Vol.~\bibinfo {volume} {2}\ (\bibinfo  {publisher} {Oxford
		university press},\ \bibinfo {year} {2008})\BibitemShut {NoStop}%
	\bibitem [{\citenamefont {Wu}\ \emph {et~al.}(2025)\citenamefont {Wu},
		\citenamefont {Fan},\ and\ \citenamefont {Chen}}]{q38s-k2tq}%
	\BibitemOpen
	\bibfield  {author} {\bibinfo {author} {\bibfnamefont {H.}~\bibnamefont
			{Wu}}, \bibinfo {author} {\bibfnamefont {X.}~\bibnamefont {Fan}},\ and\
		\bibinfo {author} {\bibfnamefont {L.}~\bibnamefont {Chen}},\ }\bibfield
	{title} {\bibinfo {title} {How gravitational waves change photon orbital
			angular momentum quantum states},\ }\href {https://doi.org/10.1103/q38s-k2tq}
	{\bibfield  {journal} {\bibinfo  {journal} {Phys. Rev. D}\ }\textbf {\bibinfo
			{volume} {112}},\ \bibinfo {pages} {124002} (\bibinfo {year}
		{2025})}\BibitemShut {NoStop}%
	\bibitem [{\citenamefont {Allen}\ \emph {et~al.}(1992)\citenamefont {Allen},
		\citenamefont {Beijersbergen}, \citenamefont {Spreeuw},\ and\ \citenamefont
		{Woerdman}}]{allen1992orbital}%
	\BibitemOpen
	\bibfield  {author} {\bibinfo {author} {\bibfnamefont {L.}~\bibnamefont
			{Allen}}, \bibinfo {author} {\bibfnamefont {M.~W.}\ \bibnamefont
			{Beijersbergen}}, \bibinfo {author} {\bibfnamefont {R.}~\bibnamefont
			{Spreeuw}},\ and\ \bibinfo {author} {\bibfnamefont {J.}~\bibnamefont
			{Woerdman}},\ }\bibfield  {title} {\bibinfo {title} {Orbital angular momentum
			of light and the transformation of laguerre-gaussian laser modes},\ }\href
	{https://doi.org/10.1103/PhysRevA.45.8185} {\bibfield  {journal} {\bibinfo
			{journal} {Phys. Rev. A}\ }\textbf {\bibinfo {volume} {45}},\ \bibinfo
		{pages} {8185} (\bibinfo {year} {1992})}\BibitemShut {NoStop}%
	\bibitem [{\citenamefont {Mair}\ \emph {et~al.}(2001)\citenamefont {Mair},
		\citenamefont {Vaziri}, \citenamefont {Weihs},\ and\ \citenamefont
		{Zeilinger}}]{mair2001entanglement}%
	\BibitemOpen
	\bibfield  {author} {\bibinfo {author} {\bibfnamefont {A.}~\bibnamefont
			{Mair}}, \bibinfo {author} {\bibfnamefont {A.}~\bibnamefont {Vaziri}},
		\bibinfo {author} {\bibfnamefont {G.}~\bibnamefont {Weihs}},\ and\ \bibinfo
		{author} {\bibfnamefont {A.}~\bibnamefont {Zeilinger}},\ }\bibfield  {title}
	{\bibinfo {title} {Entanglement of the orbital angular momentum states of
			photons},\ }\href {https://doi.org/10.1038/35085529} {\bibfield  {journal}
		{\bibinfo  {journal} {Nature}\ }\textbf {\bibinfo {volume} {412}},\ \bibinfo
		{pages} {313} (\bibinfo {year} {2001})}\BibitemShut {NoStop}%
	\bibitem [{\citenamefont {Vaziri}\ \emph {et~al.}(2002)\citenamefont {Vaziri},
		\citenamefont {Weihs},\ and\ \citenamefont
		{Zeilinger}}]{vaziri2002experimental}%
	\BibitemOpen
	\bibfield  {author} {\bibinfo {author} {\bibfnamefont {A.}~\bibnamefont
			{Vaziri}}, \bibinfo {author} {\bibfnamefont {G.}~\bibnamefont {Weihs}},\ and\
		\bibinfo {author} {\bibfnamefont {A.}~\bibnamefont {Zeilinger}},\ }\bibfield
	{title} {\bibinfo {title} {Experimental two-photon, three-dimensional
			entanglement for quantum communication},\ }\href
	{https://doi.org/10.1103/PhysRevLett.89.240401} {\bibfield  {journal}
		{\bibinfo  {journal} {Phys. Rev. Lett.}\ }\textbf {\bibinfo {volume} {89}},\
		\bibinfo {pages} {240401} (\bibinfo {year} {2002})}\BibitemShut {NoStop}%
	\bibitem [{\citenamefont {Vaziri}\ \emph {et~al.}(2003)\citenamefont {Vaziri},
		\citenamefont {Pan}, \citenamefont {Jennewein}, \citenamefont {Weihs},\ and\
		\citenamefont {Zeilinger}}]{vaziri2003concentration}%
	\BibitemOpen
	\bibfield  {author} {\bibinfo {author} {\bibfnamefont {A.}~\bibnamefont
			{Vaziri}}, \bibinfo {author} {\bibfnamefont {J.-W.}\ \bibnamefont {Pan}},
		\bibinfo {author} {\bibfnamefont {T.}~\bibnamefont {Jennewein}}, \bibinfo
		{author} {\bibfnamefont {G.}~\bibnamefont {Weihs}},\ and\ \bibinfo {author}
		{\bibfnamefont {A.}~\bibnamefont {Zeilinger}},\ }\bibfield  {title} {\bibinfo
		{title} {Concentration of higher dimensional entanglement: qutrits of photon
			orbital angular momentum},\ }\href
	{https://doi.org/10.1103/PhysRevLett.91.227902} {\bibfield  {journal}
		{\bibinfo  {journal} {Phys. Rev. Lett.}\ }\textbf {\bibinfo {volume} {91}},\
		\bibinfo {pages} {227902} (\bibinfo {year} {2003})}\BibitemShut {NoStop}%
	\bibitem [{\citenamefont {Molina-Terriza}\ \emph {et~al.}(2007)\citenamefont
		{Molina-Terriza}, \citenamefont {Torres},\ and\ \citenamefont
		{Torner}}]{molina2007twisted}%
	\BibitemOpen
	\bibfield  {author} {\bibinfo {author} {\bibfnamefont {G.}~\bibnamefont
			{Molina-Terriza}}, \bibinfo {author} {\bibfnamefont {J.~P.}\ \bibnamefont
			{Torres}},\ and\ \bibinfo {author} {\bibfnamefont {L.}~\bibnamefont
			{Torner}},\ }\bibfield  {title} {\bibinfo {title} {Twisted photons},\ }\href
	{https://doi.org/10.1038/nphys607} {\bibfield  {journal} {\bibinfo  {journal}
			{Nature Phys.}\ }\textbf {\bibinfo {volume} {3}},\ \bibinfo {pages} {305}
		(\bibinfo {year} {2007})}\BibitemShut {NoStop}%
	\bibitem [{\citenamefont {Chelkowski}\ \emph {et~al.}(2009)\citenamefont
		{Chelkowski}, \citenamefont {Hild},\ and\ \citenamefont
		{Freise}}]{PhysRevD.79.122002}%
	\BibitemOpen
	\bibfield  {author} {\bibinfo {author} {\bibfnamefont {S.}~\bibnamefont
			{Chelkowski}}, \bibinfo {author} {\bibfnamefont {S.}~\bibnamefont {Hild}},\
		and\ \bibinfo {author} {\bibfnamefont {A.}~\bibnamefont {Freise}},\
	}\bibfield  {title} {\bibinfo {title} {Prospects of higher-order
			laguerre-gauss modes in future gravitational wave detectors},\ }\href
	{https://doi.org/10.1103/PhysRevD.79.122002} {\bibfield  {journal} {\bibinfo
			{journal} {Phys. Rev. D}\ }\textbf {\bibinfo {volume} {79}},\ \bibinfo
		{pages} {122002} (\bibinfo {year} {2009})}\BibitemShut {NoStop}%
	\bibitem [{\citenamefont {Granata}\ \emph {et~al.}(2010)\citenamefont
		{Granata}, \citenamefont {Buy}, \citenamefont {Ward},\ and\ \citenamefont
		{Barsuglia}}]{PhysRevLett.105.231102}%
	\BibitemOpen
	\bibfield  {author} {\bibinfo {author} {\bibfnamefont {M.}~\bibnamefont
			{Granata}}, \bibinfo {author} {\bibfnamefont {C.}~\bibnamefont {Buy}},
		\bibinfo {author} {\bibfnamefont {R.}~\bibnamefont {Ward}},\ and\ \bibinfo
		{author} {\bibfnamefont {M.}~\bibnamefont {Barsuglia}},\ }\bibfield  {title}
	{\bibinfo {title} {Higher-order laguerre-gauss mode generation and
			interferometry for gravitational wave detectors},\ }\href
	{https://doi.org/10.1103/PhysRevLett.105.231102} {\bibfield  {journal}
		{\bibinfo  {journal} {Phys. Rev. Lett.}\ }\textbf {\bibinfo {volume} {105}},\
		\bibinfo {pages} {231102} (\bibinfo {year} {2010})}\BibitemShut {NoStop}%
	\bibitem [{\citenamefont {Wu}\ \emph {et~al.}(2022{\natexlab{a}})\citenamefont
		{Wu}, \citenamefont {Zhu},\ and\ \citenamefont {Feng}}]{wu2022testing}%
	\BibitemOpen
	\bibfield  {author} {\bibinfo {author} {\bibfnamefont {Q.}~\bibnamefont
			{Wu}}, \bibinfo {author} {\bibfnamefont {W.}~\bibnamefont {Zhu}},\ and\
		\bibinfo {author} {\bibfnamefont {L.}~\bibnamefont {Feng}},\ }\bibfield
	{title} {\bibinfo {title} {Testing the wave-particle duality of gravitational
			wave using the spin-orbital-hall effect of structured light},\ }\href
	{https://doi.org/10.3390/universe8100535} {\bibfield  {journal} {\bibinfo
			{journal} {Universe}\ }\textbf {\bibinfo {volume} {8}},\ \bibinfo {pages}
		{535} (\bibinfo {year} {2022}{\natexlab{a}})}\BibitemShut {NoStop}%
	\bibitem [{\citenamefont {Wu}\ \emph {et~al.}(2022{\natexlab{b}})\citenamefont
		{Wu}, \citenamefont {Fan},\ and\ \citenamefont {Chen}}]{PhysRevD.106.045023}%
	\BibitemOpen
	\bibfield  {author} {\bibinfo {author} {\bibfnamefont {H.}~\bibnamefont
			{Wu}}, \bibinfo {author} {\bibfnamefont {X.}~\bibnamefont {Fan}},\ and\
		\bibinfo {author} {\bibfnamefont {L.}~\bibnamefont {Chen}},\ }\bibfield
	{title} {\bibinfo {title} {How gravitational fluctuations degrade the
			high-dimensional spatial entanglement},\ }\href
	{https://doi.org/10.1103/PhysRevD.106.045023} {\bibfield  {journal} {\bibinfo
			{journal} {Phys. Rev. D}\ }\textbf {\bibinfo {volume} {106}},\ \bibinfo
		{pages} {045023} (\bibinfo {year} {2022}{\natexlab{b}})}\BibitemShut
	{NoStop}%
	\bibitem [{\citenamefont {Gross}\ \emph {et~al.}(2010)\citenamefont {Gross},
		\citenamefont {Liu}, \citenamefont {Flammia}, \citenamefont {Becker},\ and\
		\citenamefont {Eisert}}]{PhysRevLett.105.150401}%
	\BibitemOpen
	\bibfield  {author} {\bibinfo {author} {\bibfnamefont {D.}~\bibnamefont
			{Gross}}, \bibinfo {author} {\bibfnamefont {Y.-K.}\ \bibnamefont {Liu}},
		\bibinfo {author} {\bibfnamefont {S.~T.}\ \bibnamefont {Flammia}}, \bibinfo
		{author} {\bibfnamefont {S.}~\bibnamefont {Becker}},\ and\ \bibinfo {author}
		{\bibfnamefont {J.}~\bibnamefont {Eisert}},\ }\bibfield  {title} {\bibinfo
		{title} {Quantum state tomography via compressed sensing},\ }\href
	{https://doi.org/10.1103/PhysRevLett.105.150401} {\bibfield  {journal}
		{\bibinfo  {journal} {Phys. Rev. Lett.}\ }\textbf {\bibinfo {volume} {105}},\
		\bibinfo {pages} {150401} (\bibinfo {year} {2010})}\BibitemShut {NoStop}%
	\bibitem [{\citenamefont {Schreiber}\ \emph {et~al.}(2025)\citenamefont
		{Schreiber}, \citenamefont {Eisert},\ and\ \citenamefont
		{Meyer}}]{PRXQuantum.6.020346}%
	\BibitemOpen
	\bibfield  {author} {\bibinfo {author} {\bibfnamefont {F.~J.}\ \bibnamefont
			{Schreiber}}, \bibinfo {author} {\bibfnamefont {J.}~\bibnamefont {Eisert}},\
		and\ \bibinfo {author} {\bibfnamefont {J.~J.}\ \bibnamefont {Meyer}},\
	}\bibfield  {title} {\bibinfo {title} {Tomography of parametrized quantum
			states},\ }\href {https://doi.org/10.1103/PRXQuantum.6.020346} {\bibfield
		{journal} {\bibinfo  {journal} {PRX Quantum}\ }\textbf {\bibinfo {volume}
			{6}},\ \bibinfo {pages} {020346} (\bibinfo {year} {2025})}\BibitemShut
	{NoStop}%
	\bibitem [{\citenamefont {Tonner}\ \emph {et~al.}(2026)\citenamefont {Tonner},
		\citenamefont {Tran},\ and\ \citenamefont
		{Kueng}}]{tonner2026parametricquantumstatetomography}%
	\BibitemOpen
	\bibfield  {author} {\bibinfo {author} {\bibfnamefont {S.}~\bibnamefont
			{Tonner}}, \bibinfo {author} {\bibfnamefont {V.~T.}\ \bibnamefont {Tran}},\
		and\ \bibinfo {author} {\bibfnamefont {R.}~\bibnamefont {Kueng}},\ }\href
	{https://arxiv.org/abs/2601.20950} {\bibinfo {title} {Parametric quantum
			state tomography with hyperrbms}} (\bibinfo {year} {2026}),\ \Eprint
	{https://arxiv.org/abs/2601.20950} {arXiv:2601.20950 [quant-ph]} \BibitemShut
	{NoStop}%
	\bibitem [{sup()}]{suppleTomography}%
	\BibitemOpen
	\href@noop {} {}\bibinfo {note} {The derivation for OAM transition amplitudes
		induced by GWs and state evolution in the Mach-Zehnder interferometer are
		given in Supplemental Material at [URL will be inserted by
		publisher].}\BibitemShut {Stop}%
	\bibitem [{\citenamefont {Arlt}\ and\ \citenamefont
		{Dholakia}(2000)}]{ARLT2000297}%
	\BibitemOpen
	\bibfield  {author} {\bibinfo {author} {\bibfnamefont {J.}~\bibnamefont
			{Arlt}}\ and\ \bibinfo {author} {\bibfnamefont {K.}~\bibnamefont
			{Dholakia}},\ }\bibfield  {title} {\bibinfo {title} {Generation of high-order
			bessel beams by use of an axicon},\ }\href
	{https://doi.org/https://doi.org/10.1016/S0030-4018(00)00572-1} {\bibfield
		{journal} {\bibinfo  {journal} {Opt. Commun.}\ }\textbf {\bibinfo {volume}
			{177}},\ \bibinfo {pages} {297} (\bibinfo {year} {2000})}\BibitemShut
	{NoStop}%
	\bibitem [{\citenamefont {Durnin}\ \emph {et~al.}(1987)\citenamefont {Durnin},
		\citenamefont {Miceli},\ and\ \citenamefont {Eberly}}]{PhysRevLett.58.1499}%
	\BibitemOpen
	\bibfield  {author} {\bibinfo {author} {\bibfnamefont {J.}~\bibnamefont
			{Durnin}}, \bibinfo {author} {\bibfnamefont {J.~J.}\ \bibnamefont {Miceli}},\
		and\ \bibinfo {author} {\bibfnamefont {J.~H.}\ \bibnamefont {Eberly}},\
	}\bibfield  {title} {\bibinfo {title} {Diffraction-free beams},\ }\href
	{https://doi.org/10.1103/PhysRevLett.58.1499} {\bibfield  {journal} {\bibinfo
			{journal} {Phys. Rev. Lett.}\ }\textbf {\bibinfo {volume} {58}},\ \bibinfo
		{pages} {1499} (\bibinfo {year} {1987})}\BibitemShut {NoStop}%
	\bibitem [{\citenamefont {McLaren}\ \emph {et~al.}(2013)\citenamefont
		{McLaren}, \citenamefont {Romero}, \citenamefont {Padgett}, \citenamefont
		{Roux},\ and\ \citenamefont {Forbes}}]{PhysRevA.88.033818}%
	\BibitemOpen
	\bibfield  {author} {\bibinfo {author} {\bibfnamefont {M.}~\bibnamefont
			{McLaren}}, \bibinfo {author} {\bibfnamefont {J.}~\bibnamefont {Romero}},
		\bibinfo {author} {\bibfnamefont {M.~J.}\ \bibnamefont {Padgett}}, \bibinfo
		{author} {\bibfnamefont {F.~S.}\ \bibnamefont {Roux}},\ and\ \bibinfo
		{author} {\bibfnamefont {A.}~\bibnamefont {Forbes}},\ }\bibfield  {title}
	{\bibinfo {title} {Two-photon optics of bessel-gaussian modes},\ }\href
	{https://doi.org/10.1103/PhysRevA.88.033818} {\bibfield  {journal} {\bibinfo
			{journal} {Phys. Rev. A}\ }\textbf {\bibinfo {volume} {88}},\ \bibinfo
		{pages} {033818} (\bibinfo {year} {2013})}\BibitemShut {NoStop}%
	\bibitem [{\citenamefont {Kuntz}\ \emph {et~al.}(2009)\citenamefont {Kuntz},
		\citenamefont {Braverman}, \citenamefont {Youn}, \citenamefont {Lobino},
		\citenamefont {Pessina},\ and\ \citenamefont {Lvovsky}}]{PhysRevA.79.043802}%
	\BibitemOpen
	\bibfield  {author} {\bibinfo {author} {\bibfnamefont {K.~B.}\ \bibnamefont
			{Kuntz}}, \bibinfo {author} {\bibfnamefont {B.}~\bibnamefont {Braverman}},
		\bibinfo {author} {\bibfnamefont {S.~H.}\ \bibnamefont {Youn}}, \bibinfo
		{author} {\bibfnamefont {M.}~\bibnamefont {Lobino}}, \bibinfo {author}
		{\bibfnamefont {E.~M.}\ \bibnamefont {Pessina}},\ and\ \bibinfo {author}
		{\bibfnamefont {A.~I.}\ \bibnamefont {Lvovsky}},\ }\bibfield  {title}
	{\bibinfo {title} {Spatial and temporal characterization of a bessel beam
			produced using a conical mirror},\ }\href
	{https://doi.org/10.1103/PhysRevA.79.043802} {\bibfield  {journal} {\bibinfo
			{journal} {Phys. Rev. A}\ }\textbf {\bibinfo {volume} {79}},\ \bibinfo
		{pages} {043802} (\bibinfo {year} {2009})}\BibitemShut {NoStop}%
	\bibitem [{\citenamefont {Collaboration}\ \emph {et~al.}(2015)\citenamefont
		{Collaboration}, \citenamefont {Aasi}, \citenamefont {Abbott} \emph
		{et~al.}}]{Aasi_2015}%
	\BibitemOpen
	\bibfield  {author} {\bibinfo {author} {\bibfnamefont {T.~L.~S.}\
			\bibnamefont {Collaboration}}, \bibinfo {author} {\bibfnamefont
			{J.}~\bibnamefont {Aasi}}, \bibinfo {author} {\bibfnamefont {B.~P.}\
			\bibnamefont {Abbott}}, \emph {et~al.},\ }\bibfield  {title} {\bibinfo
		{title} {Advanced ligo},\ }\href
	{https://doi.org/10.1088/0264-9381/32/7/074001} {\bibfield  {journal}
		{\bibinfo  {journal} {Classical and Quantum Gravity}\ }\textbf {\bibinfo
			{volume} {32}},\ \bibinfo {pages} {074001} (\bibinfo {year}
		{2015})}\BibitemShut {NoStop}%
	\bibitem [{\citenamefont {Leach}\ \emph {et~al.}(2002)\citenamefont {Leach},
		\citenamefont {Padgett}, \citenamefont {Barnett}, \citenamefont
		{Franke-Arnold},\ and\ \citenamefont {Courtial}}]{leach2002measuring}%
	\BibitemOpen
	\bibfield  {author} {\bibinfo {author} {\bibfnamefont {J.}~\bibnamefont
			{Leach}}, \bibinfo {author} {\bibfnamefont {M.~J.}\ \bibnamefont {Padgett}},
		\bibinfo {author} {\bibfnamefont {S.~M.}\ \bibnamefont {Barnett}}, \bibinfo
		{author} {\bibfnamefont {S.}~\bibnamefont {Franke-Arnold}},\ and\ \bibinfo
		{author} {\bibfnamefont {J.}~\bibnamefont {Courtial}},\ }\bibfield  {title}
	{\bibinfo {title} {Measuring the orbital angular momentum of a single
			photon},\ }\href {https://doi.org/10.1103/PhysRevLett.88.257901} {\bibfield
		{journal} {\bibinfo  {journal} {Phys. Rev. Lett.}\ }\textbf {\bibinfo
			{volume} {88}},\ \bibinfo {pages} {257901} (\bibinfo {year}
		{2002})}\BibitemShut {NoStop}%
	\bibitem [{\citenamefont {Leach}\ \emph {et~al.}(2004)\citenamefont {Leach},
		\citenamefont {Courtial}, \citenamefont {Skeldon}, \citenamefont {Barnett},
		\citenamefont {Franke-Arnold},\ and\ \citenamefont
		{Padgett}}]{PhysRevLett.92.013601}%
	\BibitemOpen
	\bibfield  {author} {\bibinfo {author} {\bibfnamefont {J.}~\bibnamefont
			{Leach}}, \bibinfo {author} {\bibfnamefont {J.}~\bibnamefont {Courtial}},
		\bibinfo {author} {\bibfnamefont {K.}~\bibnamefont {Skeldon}}, \bibinfo
		{author} {\bibfnamefont {S.~M.}\ \bibnamefont {Barnett}}, \bibinfo {author}
		{\bibfnamefont {S.}~\bibnamefont {Franke-Arnold}},\ and\ \bibinfo {author}
		{\bibfnamefont {M.~J.}\ \bibnamefont {Padgett}},\ }\bibfield  {title}
	{\bibinfo {title} {Interferometric methods to measure orbital and spin, or
			the total angular momentum of a single photon},\ }\href
	{https://doi.org/10.1103/PhysRevLett.92.013601} {\bibfield  {journal}
		{\bibinfo  {journal} {Phys. Rev. Lett.}\ }\textbf {\bibinfo {volume} {92}},\
		\bibinfo {pages} {013601} (\bibinfo {year} {2004})}\BibitemShut {NoStop}%
	\bibitem [{\citenamefont {Jha}\ \emph {et~al.}(2011)\citenamefont {Jha},
		\citenamefont {Agarwal},\ and\ \citenamefont {Boyd}}]{PhysRevA.83.053829}%
	\BibitemOpen
	\bibfield  {author} {\bibinfo {author} {\bibfnamefont {A.~K.}\ \bibnamefont
			{Jha}}, \bibinfo {author} {\bibfnamefont {G.~S.}\ \bibnamefont {Agarwal}},\
		and\ \bibinfo {author} {\bibfnamefont {R.~W.}\ \bibnamefont {Boyd}},\
	}\bibfield  {title} {\bibinfo {title} {Supersensitive measurement of angular
			displacements using entangled photons},\ }\href
	{https://doi.org/10.1103/PhysRevA.83.053829} {\bibfield  {journal} {\bibinfo
			{journal} {Phys. Rev. A}\ }\textbf {\bibinfo {volume} {83}},\ \bibinfo
		{pages} {053829} (\bibinfo {year} {2011})}\BibitemShut {NoStop}%
	\bibitem [{\citenamefont {Maggiore}(2007)}]{9780198570745.001.0001}%
	\BibitemOpen
	\bibfield  {author} {\bibinfo {author} {\bibfnamefont {M.}~\bibnamefont
			{Maggiore}},\ }\href
	{https://doi.org/10.1093/acprof:oso/9780198570745.001.0001} {\emph {\bibinfo
			{title} {Gravitational Waves: Volume 1: Theory and Experiments}}}\ (\bibinfo
	{publisher} {Oxford University Press},\ \bibinfo {address} {Oxford},\
	\bibinfo {year} {2007})\BibitemShut {NoStop}%
	\bibitem [{\citenamefont {Liu}\ \emph {et~al.}(2026)\citenamefont {Liu},
		\citenamefont {Chen}, \citenamefont {Zhao}, \citenamefont {Wang},
		\citenamefont {Liu},\ and\ \citenamefont {Wang}}]{Liu:26}%
	\BibitemOpen
	\bibfield  {author} {\bibinfo {author} {\bibfnamefont {Q.}~\bibnamefont
			{Liu}}, \bibinfo {author} {\bibfnamefont {Y.}~\bibnamefont {Chen}}, \bibinfo
		{author} {\bibfnamefont {Z.}~\bibnamefont {Zhao}}, \bibinfo {author}
		{\bibfnamefont {Z.}~\bibnamefont {Wang}}, \bibinfo {author} {\bibfnamefont
			{Y.}~\bibnamefont {Liu}},\ and\ \bibinfo {author} {\bibfnamefont
			{P.}~\bibnamefont {Wang}},\ }\bibfield  {title} {\bibinfo {title}
		{Propagation of gaussian and bessel--gaussian beams through disturbances: an
			analysis of ranging efficiency},\ }\href {https://doi.org/10.1364/AO.578388}
	{\bibfield  {journal} {\bibinfo  {journal} {Appl. Opt.}\ }\textbf {\bibinfo
			{volume} {65}},\ \bibinfo {pages} {487} (\bibinfo {year} {2026})}\BibitemShut
	{NoStop}%
	\bibitem [{\citenamefont {Misner}\ \emph {et~al.}(1973)\citenamefont {Misner},
		\citenamefont {Thorne},\ and\ \citenamefont
		{Wheeler}}]{misner1973gravitation}%
	\BibitemOpen
	\bibfield  {author} {\bibinfo {author} {\bibfnamefont {C.~W.}\ \bibnamefont
			{Misner}}, \bibinfo {author} {\bibfnamefont {K.~S.}\ \bibnamefont {Thorne}},\
		and\ \bibinfo {author} {\bibfnamefont {J.~A.}\ \bibnamefont {Wheeler}},\
	}\href@noop {} {\emph {\bibinfo {title} {Gravitation}}}\ (\bibinfo
	{publisher} {Macmillan},\ \bibinfo {address} {London},\ \bibinfo {year}
	{1973})\BibitemShut {NoStop}%
	\bibitem [{\citenamefont {Painchaud}\ \emph {et~al.}(2009)\citenamefont
		{Painchaud}, \citenamefont {Poulin}, \citenamefont {Morin},\ and\
		\citenamefont {T\^{e}tu}}]{Painchaud:09}%
	\BibitemOpen
	\bibfield  {author} {\bibinfo {author} {\bibfnamefont {Y.}~\bibnamefont
			{Painchaud}}, \bibinfo {author} {\bibfnamefont {M.}~\bibnamefont {Poulin}},
		\bibinfo {author} {\bibfnamefont {M.}~\bibnamefont {Morin}},\ and\ \bibinfo
		{author} {\bibfnamefont {M.}~\bibnamefont {T\^{e}tu}},\ }\bibfield  {title}
	{\bibinfo {title} {Performance of balanced detection in a coherent
			receiver},\ }\href {https://doi.org/10.1364/OE.17.003659} {\bibfield
		{journal} {\bibinfo  {journal} {Opt. Express}\ }\textbf {\bibinfo {volume}
			{17}},\ \bibinfo {pages} {3659} (\bibinfo {year} {2009})}\BibitemShut
	{NoStop}%
	\bibitem [{\citenamefont {Fricke}\ \emph {et~al.}(2012)\citenamefont {Fricke},
		\citenamefont {Smith-Lefebvre}, \citenamefont {Abbott}, \citenamefont
		{Adhikari}, \citenamefont {Dooley}, \citenamefont {Evans}, \citenamefont
		{Fritschel}, \citenamefont {Frolov}, \citenamefont {Kawabe}, \citenamefont
		{Kissel} \emph {et~al.}}]{fricke2012dc}%
	\BibitemOpen
	\bibfield  {author} {\bibinfo {author} {\bibfnamefont {T.~T.}\ \bibnamefont
			{Fricke}}, \bibinfo {author} {\bibfnamefont {N.~D.}\ \bibnamefont
			{Smith-Lefebvre}}, \bibinfo {author} {\bibfnamefont {R.}~\bibnamefont
			{Abbott}}, \bibinfo {author} {\bibfnamefont {R.}~\bibnamefont {Adhikari}},
		\bibinfo {author} {\bibfnamefont {K.~L.}\ \bibnamefont {Dooley}}, \bibinfo
		{author} {\bibfnamefont {M.}~\bibnamefont {Evans}}, \bibinfo {author}
		{\bibfnamefont {P.}~\bibnamefont {Fritschel}}, \bibinfo {author}
		{\bibfnamefont {V.~V.}\ \bibnamefont {Frolov}}, \bibinfo {author}
		{\bibfnamefont {K.}~\bibnamefont {Kawabe}}, \bibinfo {author} {\bibfnamefont
			{J.~S.}\ \bibnamefont {Kissel}}, \emph {et~al.},\ }\bibfield  {title}
	{\bibinfo {title} {Dc readout experiment in enhanced ligo},\ }\href
	{https://doi.org/10.1088/0264-9381/29/6/065005} {\bibfield  {journal}
		{\bibinfo  {journal} {Classical and Quantum Gravity}\ }\textbf {\bibinfo
			{volume} {29}},\ \bibinfo {pages} {065005} (\bibinfo {year}
		{2012})}\BibitemShut {NoStop}%
	\bibitem [{\citenamefont {Zhang}\ \emph {et~al.}(2021)\citenamefont {Zhang},
		\citenamefont {Jones}, \citenamefont {Smetana}, \citenamefont {Miao},
		\citenamefont {Martynov}, \citenamefont {Freise},\ and\ \citenamefont
		{Ballmer}}]{PhysRevLett.126.221301}%
	\BibitemOpen
	\bibfield  {author} {\bibinfo {author} {\bibfnamefont {T.}~\bibnamefont
			{Zhang}}, \bibinfo {author} {\bibfnamefont {P.}~\bibnamefont {Jones}},
		\bibinfo {author} {\bibfnamefont {J.~c.~v.}\ \bibnamefont {Smetana}},
		\bibinfo {author} {\bibfnamefont {H.}~\bibnamefont {Miao}}, \bibinfo {author}
		{\bibfnamefont {D.}~\bibnamefont {Martynov}}, \bibinfo {author}
		{\bibfnamefont {A.}~\bibnamefont {Freise}},\ and\ \bibinfo {author}
		{\bibfnamefont {S.~W.}\ \bibnamefont {Ballmer}},\ }\bibfield  {title}
	{\bibinfo {title} {Two-carrier scheme: Evading the 3 db quantum penalty of
			heterodyne readout in gravitational-wave detectors},\ }\href
	{https://doi.org/10.1103/PhysRevLett.126.221301} {\bibfield  {journal}
		{\bibinfo  {journal} {Phys. Rev. Lett.}\ }\textbf {\bibinfo {volume} {126}},\
		\bibinfo {pages} {221301} (\bibinfo {year} {2021})}\BibitemShut {NoStop}%
	\bibitem [{\citenamefont {Martynov}\ \emph {et~al.}(2016)\citenamefont
		{Martynov}, \citenamefont {Hall}, \citenamefont {Abbott} \emph
		{et~al.}}]{PhysRevD.93.112004}%
	\BibitemOpen
	\bibfield  {author} {\bibinfo {author} {\bibfnamefont {D.~V.}\ \bibnamefont
			{Martynov}}, \bibinfo {author} {\bibfnamefont {E.~D.}\ \bibnamefont {Hall}},
		\bibinfo {author} {\bibfnamefont {B.~P.}\ \bibnamefont {Abbott}}, \emph
		{et~al.},\ }\bibfield  {title} {\bibinfo {title} {Sensitivity of the advanced
			ligo detectors at the beginning of gravitational wave astronomy},\ }\href
	{https://doi.org/10.1103/PhysRevD.93.112004} {\bibfield  {journal} {\bibinfo
			{journal} {Phys. Rev. D}\ }\textbf {\bibinfo {volume} {93}},\ \bibinfo
		{pages} {112004} (\bibinfo {year} {2016})}\BibitemShut {NoStop}%
	\bibitem [{\citenamefont {Abbott}\ \emph
		{et~al.}(2016{\natexlab{b}})\citenamefont {Abbott}, \citenamefont {Abbott},
		\citenamefont {Abbott} \emph {et~al.}}]{PhysRevLett.116.131103}%
	\BibitemOpen
	\bibfield  {author} {\bibinfo {author} {\bibfnamefont {B.~P.}\ \bibnamefont
			{Abbott}}, \bibinfo {author} {\bibfnamefont {R.}~\bibnamefont {Abbott}},
		\bibinfo {author} {\bibfnamefont {T.~D.}\ \bibnamefont {Abbott}}, \emph
		{et~al.} (\bibinfo {collaboration} {LIGO Scientific Collaboration and Virgo
			Collaboration}),\ }\bibfield  {title} {\bibinfo {title} {Gw150914: The
			advanced ligo detectors in the era of first discoveries},\ }\href
	{https://doi.org/10.1103/PhysRevLett.116.131103} {\bibfield  {journal}
		{\bibinfo  {journal} {Phys. Rev. Lett.}\ }\textbf {\bibinfo {volume} {116}},\
		\bibinfo {pages} {131103} (\bibinfo {year} {2016}{\natexlab{b}})}\BibitemShut
	{NoStop}%
	\bibitem [{\citenamefont {Hall}\ \emph {et~al.}(2021)\citenamefont {Hall},
		\citenamefont {Kuns}, \citenamefont {Smith}, \citenamefont {Bai},
		\citenamefont {Wipf}, \citenamefont {Biscans}, \citenamefont {Adhikari},
		\citenamefont {Arai}, \citenamefont {Ballmer}, \citenamefont {Barsotti},
		\citenamefont {Chen}, \citenamefont {Evans}, \citenamefont {Fritschel},
		\citenamefont {Harms}, \citenamefont {Kamai}, \citenamefont {Rollins},
		\citenamefont {Shoemaker}, \citenamefont {Slagmolen}, \citenamefont {Weiss},\
		and\ \citenamefont {Yamamoto}}]{PhysRevD.103.122004}%
	\BibitemOpen
	\bibfield  {author} {\bibinfo {author} {\bibfnamefont {E.~D.}\ \bibnamefont
			{Hall}}, \bibinfo {author} {\bibfnamefont {K.}~\bibnamefont {Kuns}}, \bibinfo
		{author} {\bibfnamefont {J.~R.}\ \bibnamefont {Smith}}, \bibinfo {author}
		{\bibfnamefont {Y.}~\bibnamefont {Bai}}, \bibinfo {author} {\bibfnamefont
			{C.}~\bibnamefont {Wipf}}, \bibinfo {author} {\bibfnamefont {S.}~\bibnamefont
			{Biscans}}, \bibinfo {author} {\bibfnamefont {R.~X.}\ \bibnamefont
			{Adhikari}}, \bibinfo {author} {\bibfnamefont {K.}~\bibnamefont {Arai}},
		\bibinfo {author} {\bibfnamefont {S.}~\bibnamefont {Ballmer}}, \bibinfo
		{author} {\bibfnamefont {L.}~\bibnamefont {Barsotti}}, \bibinfo {author}
		{\bibfnamefont {Y.}~\bibnamefont {Chen}}, \bibinfo {author} {\bibfnamefont
			{M.}~\bibnamefont {Evans}}, \bibinfo {author} {\bibfnamefont
			{P.}~\bibnamefont {Fritschel}}, \bibinfo {author} {\bibfnamefont
			{J.}~\bibnamefont {Harms}}, \bibinfo {author} {\bibfnamefont
			{B.}~\bibnamefont {Kamai}}, \bibinfo {author} {\bibfnamefont {J.~G.}\
			\bibnamefont {Rollins}}, \bibinfo {author} {\bibfnamefont {D.}~\bibnamefont
			{Shoemaker}}, \bibinfo {author} {\bibfnamefont {B.~J.~J.}\ \bibnamefont
			{Slagmolen}}, \bibinfo {author} {\bibfnamefont {R.}~\bibnamefont {Weiss}},\
		and\ \bibinfo {author} {\bibfnamefont {H.}~\bibnamefont {Yamamoto}},\
	}\bibfield  {title} {\bibinfo {title} {Gravitational-wave physics with cosmic
			explorer: Limits to low-frequency sensitivity},\ }\href
	{https://doi.org/10.1103/PhysRevD.103.122004} {\bibfield  {journal} {\bibinfo
			{journal} {Phys. Rev. D}\ }\textbf {\bibinfo {volume} {103}},\ \bibinfo
		{pages} {122004} (\bibinfo {year} {2021})}\BibitemShut {NoStop}%
	\bibitem [{\citenamefont {Moore}\ \emph {et~al.}(2014)\citenamefont {Moore},
		\citenamefont {Cole},\ and\ \citenamefont {Berry}}]{moore2014gravitational}%
	\BibitemOpen
	\bibfield  {author} {\bibinfo {author} {\bibfnamefont {C.~J.}\ \bibnamefont
			{Moore}}, \bibinfo {author} {\bibfnamefont {R.~H.}\ \bibnamefont {Cole}},\
		and\ \bibinfo {author} {\bibfnamefont {C.~P.}\ \bibnamefont {Berry}},\
	}\bibfield  {title} {\bibinfo {title} {Gravitational-wave sensitivity
			curves},\ }\href {https://doi.org/10.1088/0264-9381/32/1/015014} {\bibfield
		{journal} {\bibinfo  {journal} {Classical and Quantum Gravity}\ }\textbf
		{\bibinfo {volume} {32}},\ \bibinfo {pages} {015014} (\bibinfo {year}
		{2014})}\BibitemShut {NoStop}%
	\bibitem [{\citenamefont {Jaranowski}\ \emph {et~al.}(1998)\citenamefont
		{Jaranowski}, \citenamefont {Kr\'olak},\ and\ \citenamefont
		{Schutz}}]{PhysRevD.58.063001}%
	\BibitemOpen
	\bibfield  {author} {\bibinfo {author} {\bibfnamefont {P.}~\bibnamefont
			{Jaranowski}}, \bibinfo {author} {\bibfnamefont {A.}~\bibnamefont
			{Kr\'olak}},\ and\ \bibinfo {author} {\bibfnamefont {B.~F.}\ \bibnamefont
			{Schutz}},\ }\bibfield  {title} {\bibinfo {title} {Data analysis of
			gravitational-wave signals from spinning neutron stars: The signal and its
			detection},\ }\href {https://doi.org/10.1103/PhysRevD.58.063001} {\bibfield
		{journal} {\bibinfo  {journal} {Phys. Rev. D}\ }\textbf {\bibinfo {volume}
			{58}},\ \bibinfo {pages} {063001} (\bibinfo {year} {1998})}\BibitemShut
	{NoStop}%
	\bibitem [{\citenamefont {Abbott}\ \emph
		{et~al.}(2016{\natexlab{c}})\citenamefont {Abbott}, \citenamefont {Abbott},
		\citenamefont {Abbott} \emph {et~al.}}]{PhysRevLett.116.241102}%
	\BibitemOpen
	\bibfield  {author} {\bibinfo {author} {\bibfnamefont {B.~P.}\ \bibnamefont
			{Abbott}}, \bibinfo {author} {\bibfnamefont {R.}~\bibnamefont {Abbott}},
		\bibinfo {author} {\bibfnamefont {T.~D.}\ \bibnamefont {Abbott}}, \emph
		{et~al.} (\bibinfo {collaboration} {LIGO Scientific Collaboration and Virgo
			Collaboration}),\ }\bibfield  {title} {\bibinfo {title} {Properties of the
			binary black hole merger gw150914},\ }\href
	{https://doi.org/10.1103/PhysRevLett.116.241102} {\bibfield  {journal}
		{\bibinfo  {journal} {Phys. Rev. Lett.}\ }\textbf {\bibinfo {volume} {116}},\
		\bibinfo {pages} {241102} (\bibinfo {year} {2016}{\natexlab{c}})}\BibitemShut
	{NoStop}%
	\bibitem [{\citenamefont {Abbott}\ \emph
		{et~al.}(2017{\natexlab{b}})\citenamefont {Abbott}, \citenamefont {Abbott},
		\citenamefont {Abbott} \emph {et~al.}}]{Abbott_2017}%
	\BibitemOpen
	\bibfield  {author} {\bibinfo {author} {\bibfnamefont {B.~P.}\ \bibnamefont
			{Abbott}}, \bibinfo {author} {\bibfnamefont {R.}~\bibnamefont {Abbott}},
		\bibinfo {author} {\bibfnamefont {T.~D.}\ \bibnamefont {Abbott}}, \emph
		{et~al.},\ }\bibfield  {title} {\bibinfo {title} {First search for
			gravitational waves from known pulsars with advanced ligo},\ }\href
	{https://doi.org/10.3847/1538-4357/aa677f} {\bibfield  {journal} {\bibinfo
			{journal} {Astrophys. J.}\ }\textbf {\bibinfo {volume} {839}},\ \bibinfo
		{pages} {12} (\bibinfo {year} {2017}{\natexlab{b}})}\BibitemShut {NoStop}%
\end{thebibliography}
\end{document}